%% file: main.tex
\documentclass[preprint]{aastex7}

\begin{document}

\title{The James Webb Space Telescope NIRSpec-PRISM Transmission Spectrum of the Super-Puff, Kepler-51d}

\correspondingauthor{Jessica E. Libby-Roberts}
\email{jer5346@psu.edu}

\author[0000-0002-2990-7613]{Jessica E. Libby-Roberts}
\affil{Department of Astronomy \& Astrophysics, 525 Davey Laboratory, The Pennsylvania State University, University Park, PA 16802, USA}
\affil{Center for Exoplanets and Habitable Worlds, 525 Davey Laboratory, The Pennsylvania State University, University Park, PA 16802, USA}
\email{jer5346@psu.edu}

\author[0000-0003-3355-1223]{Aaron Bello-Arufe}
\affiliation{Jet Propulsion Laboratory, California Institute of Technology, Pasadena, CA 91109, USA}
\email{aaron.bello.arufe@jpl.nasa.gov}

\author[0000-0002-3321-4924]{Zachory K. Berta-Thompson}
\affil{Department of Astrophysical and Planetary Sciences, University of Colorado Boulder, Boulder, CO 80309, USA}
\email{zach.bertathompson@colorado.edu}

\author[0000-0003-4835-0619]{Caleb I. Ca\~nas}
\altaffiliation{NASA Postdoctoral Fellow}
\affiliation{NASA Goddard Space Flight Center, 8800 Greenbelt Road, Greenbelt, MD 20771, USA}
\email{c.canas@nasa.gov}

\author[0000-0003-1728-8269]{Yayaati Chachan}
\affil{Department of Astronomy and Astrophysics, University of California, Santa Cruz, CA 95064, USA}
\email{yayaatichachan@gmail.com}

\author[0000-0003-2215-8485]{Renyu Hu}
\affil{Jet Propulsion Laboratory, California Institute of Technology, Pasadena, CA 91109, USA}
\affil{Division of Geological and Planetary Sciences, California Institute of Technology, Pasadena, CA 91125, USA}
\email{renyu.hu@jpl.nasa.gov}

\author[0000-0003-3800-7518]{Yui Kawashima}
\affil{Department of Astronomy, Graduate School of Science, Kyoto University, Kitashirakawa Oiwake-cho, Sakyo-ku, Kyoto 606-8502, Japan}
\email{yui.kawashima.721@gmail.com}

\author[0000-0001-8504-5862]{Catriona Murray}
\affil{Department of Astrophysical and Planetary Sciences, University of Colorado Boulder, Boulder, CO 80309, USA}
\email{catriona.murray@colorado.edu}

\author[0000-0003-3290-6758]{Kazumasa Ohno}
\affil{Division of Science, National Astronomical Observatory of Japan, 2-12-1 Osawa, Mitaka-shi 1818588 Tokyo, Japan}
\email{ohno.k.ab.715@gmail.com}

\author[0000-0002-4675-9069]{Armen Tokadjian}
\affiliation{Jet Propulsion Laboratory, California Institute of Technology, Pasadena, CA 91109, USA}
\email{armen.tokadjian@jpl.nasa.gov}

\author[0000-0001-9596-7983]{Suvrath Mahadevan}
\affil{Department of Astronomy \& Astrophysics, 525 Davey Laboratory, The Pennsylvania State University, University Park, PA 16802, USA}
\affil{Center for Exoplanets and Habitable Worlds, 525 Davey Laboratory, The Pennsylvania State University, University Park, PA 16802, USA}
\email{suvrath@psu.edu}

\author[0000-0003-1298-9699]{Kento Masuda}
\affil{Department of Earth and Space Science, Osaka University, Osaka 560-0043, Japan}
\email{kmasuda@ess.sci.osaka-u.ac.jp}

\author[0000-0003-1263-8637]{Leslie Hebb}
\affiliation{Physics Department, Hobart and William Smith Colleges, 300 Pulteney St., Geneva, NY 14456, USA}
\affiliation{Department of Astronomy and Carl Sagan Institute, Cornell University, 122 Sciences Drive, Ithaca, NY 14853, USA}
\email{hebb@hws.edu}

\author[0000-0002-4404-0456]{Caroline Morley}
\affil{Department of Astronomy, University of Texas at Austin, 2515 Speedway, Austin TX 78712, USA}
\email{cmorley@utexas.edu}

\author[0000-0002-3263-2251]{Guangwei Fu}
\affiliation{Department of Physics and Astronomy, Johns Hopkins University, Baltimore, MD, USA}
\email{guangweifu@gmail.com}

\author[0000-0002-8518-9601]{Peter Gao}
\affil{Earth and Planets Laboratory, Carnegie Institution for Science, 5241 Broad Branch Road, NW, Washington, DC 20015, USA}
\email{pgao@carnegiescience.edu}

\author[0000-0002-7352-7941]{Kevin B. Stevenson}
\affil{Johns Hopkins APL, 11100 Johns Hopkins Rd, Laurel, MD 20723, USA}
\email{kevin.stevenson@jhuapl.edu}

\begin{abstract}

Kepler-51 is a 500 Myr G dwarf hosting three ``super-puffs" and one low-mass non-transiting planet. Kepler-51d, the coolest ($T_{eq}\sim$350 K) transiting planet in this system is also one of the lowest density super-puffs known to date ($\rho_p=0.038 \pm 0.009\mathrm{~g~cm^{-3}}$). With a planetary mass of $M_p=$ 5.6 $\pm$ 1.2 M$_\oplus$ and a radius of $R_p=$ 9.32 $\pm$ 0.18 R$_\oplus$, the observed properties of this planet are not readily explained by most planet formation theories. Hypotheses explaining Kepler-51d's low density range from a substantial H/He envelope comprising $>$ 30\% its mass, a high-altitude haze layer, to a tilted ring system. To test these hypotheses, we present the NIRSpec-PRISM 0.6-5.3 $\mu$m transmission spectrum of Kepler-51d observed by the James Webb Space Telescope. We find a spectrum best-fit by a sloped line covering the entire wavelength range. Based on forward modeling and atmosphere retrievals, Kepler-51d likely possesses a low-metallicity atmosphere with high-altitude hazes of submicron particles sizes spanning pressures of 1--100 $\mu$bars. However, the spectrum could also be explained by a tilted ring with an estimated lifetime on the order of $\sim$0.1 Myr. We also investigate the stellar activity of this young Sun-like star, extracting a spot temperature significantly hotter than sunspots and spot covering fractions on the order of 0.1-10\% depending on assumed spot parameters.

\end{abstract}



\section{Introduction} 

Characterized by their low mass ($M_p<$ 30 M$_\oplus$) and extreme low density ($\rho_p<$ 0.3 g/cc), ``super-puffs" represent an unusual group of exoplanets that push the boundaries of our knowledge of planetary structure, formation, and evolution \citep{lee.and.chiang}. Unlike the more massive and significantly hotter inflated hot-Jupiters, low-mass super-puffs are cooler ($T_{eq}<$ 1000 K), indicating that their low-densities are the result of different inflation mechanisms. Using their masses, radii, ages, and temperatures, planetary structure models \citep[e.g.][]{lopez.and.fortney,thorngren.2016} suggest that these planets must maintain massive H/He atmospheres comprising $>$ 20\% of their total mass. For comparison, Uranus and Neptune have $\sim$15\% of their mass in H/He \citep{podolak.2019}.

For a small core to accrete this extreme amount of gas, \citet{lee.and.chiang} hypothesize that super-puffs could form in a region of a disk that is unusually cool and dust-free, enabling rapid accretion of H/He gas onto cores less massive than the 8--10 M$_\oplus$ cores required by the core accretion theory \citep{bodenheimer.1986.coreaccretion,pollack.1996}. \citet{chachan.dusttogas} showed that radial drift of dust creates regions at $1-10$ AU (typically beyond the water snowline) with low dust-to-gas ratios where super-earth cores could accrete substantial gaseous envelopes and turn into super-puffs. 

Most super-puffs reside in multi-planet systems with near-resonant periods. The near-resonant nature of super-puff systems is compatible with gas accretion further out, followed by gentle migration inward over time. However, whether this is an astrophysical signature of their formation or a bias of using transit timing variations (TTVs) to measure masses is still unknown. Given the duration over which they need to accrete their massive envelopes and the necessity of halting their migration inward from the star to shield them from envelope loss, super-puffs are rare in the close-in exoplanet population, with $<$ 20 super-puffs known to date.

Recent work by \citet{lammers.and.winn} and \citet{liu.kepler51d.oblateness} suggests that this formation might not be as straightforward as expected. Both works measured a slow planetary rotation period for Kepler-51d of $>$ 33 hours due to the lack of oblateness observed in the James Webb Space Telescope (JWST) NIRSpec-PRISM white light curve. As planets accrete H/He from the disk, they also gain angular momentum \citep[e.g.][]{ormel.2015}. Rapid accretion of this gas for super-puffs would suggest that these planets rotate near their break-up speeds \citep[similar to our own Solar System gas giants;][]{lammers.and.winn}. However, the slow rotation of Kepler-51d indicates that either this planet accreted gas with little angular momentum transfer or an additional mechanism created a planetary spin-down effect \citep{liu.kepler51d.oblateness}. Given Kepler-51d's distance from its host star, tidally induced spin-down is unlikely \citep{murray.and.dermott,lammers.and.winn}.

Kepler-51 is a 500 Myr Sun-like star \citep[$M_\star=$ 0.96 M$_\odot$, $R_\star=$ 0.87 R$_\odot$;][]{masuda.kepler51e} hosting three \textit{transiting} Saturn-sized planets with radii between 7--10 R$_\oplus$ on orbital periods of 45, 85, and 130 days \citep[Kepler-51b, -51c, -51d, respectively;][]{steffen.2013,masuda2014}. Their near-resonance periods yield strong TTVs which enabled \citet{masuda2014} to derive mass upper limits of $<$ 10 M$_\oplus$ for all three planets, making them the lowest density exoplanets known to date (densities $<$ 0.1 g/cc).\footnote{Masses were later revised in \citet{masuda.kepler51e}, including discovery of a fourth non-transiting planets, although masses of all three transiting planets maintain an upper 3$\sigma$ limit $<$ 10 M$_\oplus$.} One possible explanation was radius inflation due to their young ages. However, using planetary structure models from \citet{lopez2013}, \citet{libbyroberts2020} demonstrated that while the three planets are expected to continue contracting overtime, they will experience minimal atmospheric loss and maintain densities $<$ 0.3 g/cc in the next 5 Gyrs. Moreover, their large orbital distances and near circular orbits suggest that these planets are unlikely to experience significant tidal heating from gravitational interaction with their host star \citep{millholland.2019,millholland.2020}.

\citet{libbyroberts2020} used the Hubble Space Telescope Wide Field Camera 3 (HST/WFC3) to characterize the atmospheres of Kepler-51b and -51d. With scale heights $>$ 2000 km, they expected to observe large absorption features between 1.1--1.7 $\mu$m. Surprisingly, they observed a flat transmission spectra of both planets. They suggested that these planets could either possess metal-rich atmospheres $>$ 300$\times$ Solar. which would cause absorption features to be undetectable due to small scale heights, or hypothesized that both planets maintained a high-altitude haze layer at pressures $<$ 10 $\mu$bars. Based on the low densities of both planets, they favored the latter hypothesis. Motivated by this result, and the fact that two other super-puffs (Kepler-79d, \citealt{chachan2020}, and HIP 41378f, \citealt{alam.hip}) also displayed featureless HST/WFC3 spectra, alternative hypotheses were proposed. For Kepler-51b specifically, \citet{wang.and.dai} suggested that the planet is experiencing a large, dusty outflow due to significant atmospheric loss causing the planet to appear larger than expected. 

\citet{ohno2021sp} investigated collision growth of dust particles in escaping atmospheres and found that the outflow could entrain abundant dust only if it formed at high altitudes, indicating that a photochemical haze is a promising candidate for the development of a large, dusty outflow.
Relevantly, \citet{gao.and.zhang} suggested that super-puffs could maintain a ``transitional haze" that is not in hydrostatic equilibrium with the rest of the planet and, instead, is detached from the bulk of the atmosphere at significantly low pressures. Assuming this haze is optically thick, the planets would appear larger in the optical than one would expect. 

As an alternative to the haze hypothesis, a tilted ring system was also proposed for super-puffs on longer orbits, such as HIP 41378f, or younger systems, such as Kepler-51, which might still be in the process of planetary and system-wide evolution \citep{piro.rings,akinsanmi.rings}. 

Transmission spectroscopy can disentangle the different hypotheses proposed for the origin of super-puffs. A high-altitude haze layer above a large H/He atmosphere would produce a flat or sloped transmission spectrum, dependent on haze properties such as monomer production rate, haze particle shapes and sizes, and haze composition \citep{kawashima2019a,chachan2020,ohno2021sp}. A dusty outflow in turn would produce a flat spectrum across the optical and near-infrared wavelengths as well as potentially inducing small discrepancies in the ingress or egress of the transit shape \citep{wang.and.dai,gao.and.zhang}. In contrast, an optically-thick tilted ring is expected to produce a flat transmission spectrum \citep{OhnoFortney22_ringspec,alam.hip}. A metal-rich atmosphere, though not supported by the planet's low density, would also produce a relatively flat spectrum due to the resulting smaller scale height. Although the previous HST/WFC3 transmission spectrum \citep{libbyroberts2020} was unable to distinguish these hypotheses given the limited wavelength coverage, follow-up observations with greater precision and wavelength coverage potentially enables us to unveil the nature of super-puffs (see Section 6.4 of \citealt{ohno2021sp} for relevant discussions).

In this paper, we present the 0.6--5.3 $\mu$m transmission spectrum of the outer transiting Kepler-51 planet, Kepler-51d, observed with the JWST NIRSpec-PRISM. The paper is structured as follows. We detail the observations and multiple data reductions performed in Section~\ref{sec:observation}. We analyze both the broadband light curve and the spectroscopic transit curves in Section~\ref{sec:analysis} while presenting the finalized transmission spectrum with forward modeling and retrievals in Section~\ref{sec:transmission_spectrum}. Section~\ref{sec:planet_discussion} discusses the implication of Kepler-51d's unusual spectrum, while Section~\ref{sec:stellar_discussion} details the stellar spectrum and its constraints on stellar activity. We conclude in Section~\ref{sec:conclusion}.

\section{Observations and Data Reduction} \label{sec:observation}

\subsection{James Webb Space Telescope Observations}

We observed a single transit of Kepler-51d (J$_\mathrm{mag}$: 13.56) with the James Webb Space Telescope (JWST) NIRSpec-PRISM \citep{birkmann.nirpsec} on 2023 June 26 UT using NIRSpec's Bright Object Time Series (BOTS)\footnote{Program GO-2571; PI J. Libby-Roberts}. We employed 12 groups per integration yielding counts $<$ 60\% the well-depth to maintain linearity throughout an integration. Each integration had an exposure time of 2.9 seconds, and we obtained 18,082 total integrations over the $\sim$14 hour observation. As the number of planned frames exceeded the maximum value allowed due to onboard storage, there is a 2.7 minute pause half-way through the observation to allow for a data download. Given the 8.5-hour transit duration, this short pause had no impact on the overall light curve. We used the NRSRAPID readout with the S1600A slit and the SUB512 subarray to minimize read time and maximize the duty cycle. There were no additional contaminating stellar spectra detected in the SUB512 observations; a feature noted by \citet{libbyroberts2020} and \citet{masuda.kepler51e} in HST/WFC3 observations.

\subsection{Data Reduction} \label{sec:reduction}

\subsubsection{Eureka! Pipeline}

We reduced the six segments of \texttt{uncal} data products using the \texttt{Eureka!} v0.10 Pipeline \citep{Bell2022}. Starting with the Stage 1 reduction, we performed the steps recommended by the \texttt{jwst pipeline}\footnote{version 1.11.4} highlighted in \citet{rustamkulov2022}. We do not skip the jump fitting as all 12 groups remain unsaturated and assumed a cosmic ray threshold of 5$\sigma$. We applied a customized bad pixel mask created by flagging pixels with a $>$ 5$\sigma$ deviation in flux compared to nearby pixels both across rows and columns. In total, this led to an additional 10 pixels masked that were not considered by the original NIRSpec-PRISM bad pixel mask generated by the \texttt{jwst pipeline}. We applied a group-level background subtraction by calculating and then subtracting a column-by-column median background value derived from the bottom 6 pixels and top 8 pixels (the stellar point-spread-function was not perfectly centered on the subarray). As the SUB512 does not include an overscan region, we performed a manual pixel-level correction based on the top and bottom 6-pixels.

\begin{figure*}
    \centering
    \includegraphics[width=.7\linewidth]{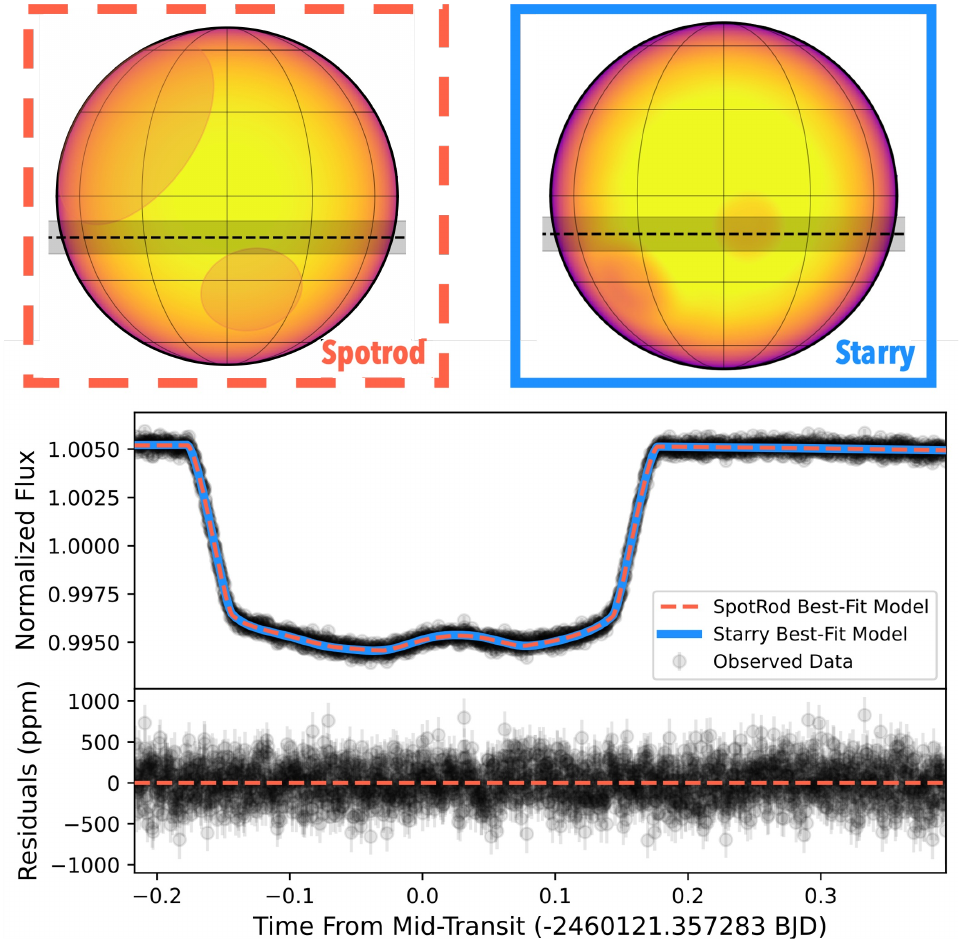}
    \caption{\textit{Top:} The best-fit \texttt{spotrod} and \texttt{starry} spot configurations represented on Kepler-51. \textit{Middle:} The JWST white light curve of Kepler-51d with flux binned from 0.6 -- 5.3 $\mu$m. Points are binned to 30 seconds for clarity. Best-fit models from the above \texttt{spotrod} and \texttt{starry} are plotted in orange and blue respectively. Regardless of the spot configuration, the white light curve models are identical. \textit{Bottom:} Plot showcasing the white light curve residuals from the \texttt{spotrod} best-fit model above. }
    \label{fig:wlc}
\end{figure*}

For Stage 2, we performed the \texttt{Eureka!} recommended steps including skipping the flat-fielding. Stage 3 extracts the final 1-dimensional stellar spectrum for each integration, assuming that all six segments are stitched together and analyzed simultaneously. To accomplish this, \texttt{Eureka!} computes the source position by fitting a Gaussian profile before summing the flux for each channel across an aperture size of 4 pixels, which minimized the overall root-mean-square (RMS) of the final light curve. A second background subtraction is then performed by calculating the median value 7 pixels away from the best-fit source position. The optimal stellar spectrum is finally extracted from the 2-dimensional data product using the optimal extraction build in \texttt{Eureka!} and assuming a 4-pixel wide aperture centered on the peak of the median Gaussian profile. We created the white light curve by summing the flux from 0.6-5.3 $\mu$m (limits imposed to maintain fluxes $>$ 500 electrons) for each integration. We also created individual spectroscopic curves from each channel yielding 408 total spectroscopic light curves with a wavelength-bin between 2 nm and 5 nm for the bluest and reddest wavelengths respectively. A rolling median boxcar of 13 integrations flagged and removed all points with $>$ 3$\sigma$ deviation. The final white light curve is plotted in Figure~\ref{fig:wlc}.

\subsubsection{ExoTic-JEDI}
We also reduced the data using \texttt{ExoTic-JEDI} \citep{alderson.exotic.wasp39b} as a pipeline comparison. We followed the prescription of the default \texttt{jwst pipeline}\footnote{version 1.12.5} for Stage 1 except that we excluded the jump and superbias subtraction step. We used the provided data-quality flags to identify bad pixels or $>$ 10$\sigma$ outliers from the median of 10 integrations. These pixels were then replaced with the median value of the surrounding 20 pixels in each row. For Stage 2, we performed the same steps as the pipeline except we skipped the flat field and extracting the 1-dimensional spectrum (which was done later). We also used \texttt{ExoTic-JEDI} to remove the 1/f noise as described in \citet{alderson.exotic.wasp39b}. The background was subtracted in each column by calculating the median value of the top and bottom 5 pixels. \texttt{ExoTic-JEDI} is set to extract an aperture defined as a certain number of full-widths-half-maximums (FWHMs) from the individual column's trace center. However, in order to compare directly with the \texttt{Eureka!} reductions, we extracted a box aperture with the same number of pixels as \texttt{Eureka!}. The final white light curve was compiled by summing the same wavelength range from 0.6-5.3 $\mu$m.

From this reduction method, we determined a comparable amount of scatter in the white light curve as the \texttt{Eureka!} reduction. We subtracted the two reduced white light curves - determining uncorrelated Gaussian residuals with a mean of 0.0 and a 1$\sigma$ spread of 340 ppm (the median absolute deviation of each light curve is $\sim$500 ppm). We therefore conclude that data reductions from both the \texttt{Eureka!} and \texttt{ExoTic-JEDI} pipelines are the same.

\section{Light Curve Analysis}\label{sec:analysis}

\subsection{White Light Curve - Spotrod}\label{sec:wlc_eureka}

The Kepler-51d transit occurred $\sim$2 hours earlier than expected due to a previously unknown non-transiting planet in the system, Kepler-51e \citep{masuda.kepler51e}. This left a pre-transit baseline of 1.2 hours in which we cropped off the initial 15 minutes due to a ramp like feature common in many NIRSpec data sets from instrument settling \citep[e.g.][]{rustamkulov2022}. We also cropped out points 5-minutes after the download pause mid-transit as the light curve showed a notable spike, likely due to the settling of the point-spread-function. This left a post-transit baseline of 5 hours and an 8.5 hour transit. A notable star spot crossing event occurs during the middle of the transit, while a tentative second smaller spot crossing event occurs 1.4 hours after the start of the transit. The spot crossings are unsurprising given the activity of the 500 Myr star and \textit{Kepler} observed multiple transits displaying spot crossing events \citep{masuda2014,libbyroberts2020}.

We analyzed both the \texttt{Eureka!} and \texttt{ExoTic-JEDI} white light curve reductions using \texttt{spotrod} \citep{spotrod} which includes a transit plus a stellar spot model. We combined \texttt{spotrod} with a second-order polynomial function to correct for out-of-transit instrumental effects. With a stellar rotation period of 8.222 days, the 14-hour JWST observation covers 14\% of the star's rotation. It is therefore possible that unocculted spots may rotate in and out during this time frame. However, a comparison of the pre-transit and post-transit stellar spectra demonstrate no detectable deviations. With no indication of significant stellar activity in this time frame, we concluded a second-order polynomial was sufficient to capture any potential stellar and instrumental variability. In total, we fit for the three parameters from the polynomial model, the transit depth, a/R$_{s}$, inclination, mid-transit time, quadratic limb darkening coefficients linearized based on \citet{kipping2013}, the x- and y-location of both spots (z-axis is assumed along the line of sight), the individual spot sizes, and a spot contrast assumed to be the same for both spots\footnote{\texttt{spotrod} assumes a spot contrast of 0 to be a completely dark spot while a contrast of 1 to be equal to the photosphere.} We also included an error scaling term, multiplied to the \texttt{Eureka!} approximated flux uncertainties. We fixed the orbital period of the planet constant to 130.18 days \citep{libbyroberts2020} and the eccentricity to zero to increase the computational efficiency. \citet{masuda.kepler51e} indicates that Kepler-51d's orbit has a slight eccentricity ($<$ 0.05) from transit timing variations (TTVs); however, this minimal eccentricity has no impact on the outcome of the transit-fit beyond increasing the computational time. 

We discovered that the spot parameters were degenerate, with multiple possible solutions in spot location, size, and contrast. In order to explore the parameter space, we opted to use dynamic nested sampling \citep{nestedsampling1} with \texttt{dynesty} \citep{dynesty1}. We used 5000 live points, with a convergence criterion of $\Delta\ln Z=0.1$. Uniform priors were assigned to the spot parameters, quadratic terms, limb darkening, and error scaling. Gaussian priors were assigned to the planetary parameters using values from \citet{libbyroberts2020}. 

The best-fit model from \texttt{Eureka!} is plotted in orange in Figure~\ref{fig:wlc} along with an image of the two-spot configuration for clarity. The best-fit parameters from both \texttt{Eureka!} and \texttt{ExoTic-JEDI} are listed in Table~\ref{tab:wlcparams} showing all parameters to be within 2$\sigma$ of each other. We perform a time-averaging test of the residuals \citep{pont.rmsnoise,kipping.notallan}, determining a Gaussian-correlation between RMS and bin-size up to 5-minutes (70 ppm) before demonstrating uncorrected correlated noise. The source of this correlated noise is unknown, though attempts to correct for it by including drift and changes in the PSF width were unsuccessful. The presence of this correlation was observed in both \texttt{Eureka!} and \texttt{ExoTic-JEDI} reductions in and out of transit. However, as discussed in Section~\ref{sec:spectroscopiclc}, the individual wavelength light curves did not demonstrate this overall structure due to their lower signal-to-noise ratio (SNR).

\input{bestfit_table}\label{tab:wlcparams}

\subsection{White Light Curve - Starry}\label{sec:wlc_starry}

Additionally, we performed a white light curve fit of the \texttt{Eureka!} reduction using the \texttt{chromatic\_fitting} package \citep{catriona_murray_chromatic_fitting_2025}. \texttt{chromatic\_fitting} utilizes \texttt{starry} for the transit and spot-crossing models combined with a second-order polynomial function for the instrumental systematics. \texttt{starry} models orbital bodies with spherical harmonic maps. Stellar spots in \texttt{starry} are generated by  using spherical harmonics to approximate a circular starspot and Gaussian-smoothing, resulting in a smooth feature on the star. This treatment differs from \texttt{spotrod}, which assumes a hard-edged circle projected onto a sphere.

For our white light curve fit, we binned the iterations to a cadence of 2 minutes. We assumed a stellar mass of 0.985\,M$_{\odot}$ \citep{masuda.kepler51e}, a stellar rotation period of 8.222 days \citep{mcquillan.rotation}, a stellar inclination of 90$^{\circ}$, an orbital period of 130.1845 days \citep{masuda.kepler51e}, and an eccentricity of 0. As in Section \ref{sec:wlc_eureka}, we fit for two spot-crossing events in the light curve. In total we fit for 18 parameters: three for the quadratic model, seven for the transit model (stellar radius, quadratic limb darkening coefficients parameterized based on \cite{kipping2013}, planet mass, planet radius, inclination, mid-transit epoch), seven for the two spots (a shared contrast, both radii, latitudes, and longitudes) and an error inflation parameter. For the stellar and planetary parameters, Gaussian priors were chosen centered on values from \citet{libbyroberts2020}. Uniform priors were chosen for both spot radii (5--30$^{\circ}$), longitudes (120--120$^{\circ}$), latitudes (0--120$^{\circ}$), and contrast (0--1)\footnote{\texttt{starry} assumes a contrast as 1 for a perfectly dark spot}. The spot latitudes and longitudes are defined here at mid-transit. Only negative latitude values were considered for the spot to avoid a double-peaked posterior (due to the symmetry above and below the transit chord).

Using \texttt{PyMC3} from within \texttt{chromatic\_fitting} to perform MCMC sampling, we ran 1000 tuning steps and 3000 draws across 2 chains. We checked that the MCMC had converged by ensuring that the Gelman-Rubin statistic for all parameters was close to 1 (all $<1.07$).
The means of the posterior distributions and 1-$\sigma$ ranges for each variable are presented in Table \ref{tab:wlcparams}. The spot latitudes and longitudes were converted to x- and y-positions on the stellar surface. We propagate the uncertainties on latitude, longitude, and stellar radius when converting to x/y by converting $\pm$1$\sigma$ values and using the smallest and largest x/y values as the 1$\sigma$ uncertainties on x and y. The best-fit white light curve model from \texttt{chromatic\_fitting} is indistinguishable by-eye from the \texttt{spotrod} model shown in Figure \ref{fig:wlc}. While the stellar spot contrast agrees well between \texttt{starry} and \texttt{spotrod}, the spot sizes and y-locations differ, likely due to the different models for spots between \texttt{starry} and \texttt{spotrod}, the prior on latitude (only negative latitudes were considered), optimized chain starting positions, and the known degeneracy between spot position and size (further discussion in Section~\ref{sec:stellar_discussion}).

\subsection{Spectroscopic Light Curves - Eureka! and ExoTic-JEDI}\label{sec:spectroscopiclc}

We created 408 spectroscopic light curves spanning 0.6--5.3 $\mu$m by summing the flux in each channel individually (``native" resolution). We modeled each spectroscopic transit curve by holding all planetary and spot parameters constant to values determined by the best-fit \texttt{Eureka!} white light curve. We fit for the wavelength dependent transit depth, quadratic limb darkening parameters, spot contrast, a linear polynomial systematic model and an error scaling term. Given the signal-to-noise (SNR) of the individual channels, we only adopted the main spot in the model -- including the second grazing spot significantly biases the overall results as the spectroscopic light curves do not possess the signal to detect this small amplitude. We also opted to use the non-re-parametrized limb darkening quadratic coefficients as recommended in \citet{coulombe.limbdark.jwst.bias}. We again used \texttt{dynesty} for each fit with 500 live points. Uniform priors were placed on all parameters with u1 and u2 bounded from -2 to 2 to enable full exploration. We also allowed for spot contrast values $>$ 1 providing us with Gaussian posteriors instead of upper limits at the redder wavelength (where spot contrast is expected to be approaching 1). We found that the residuals from all spectroscopic channel fits with wavelengths $<$ 4.5 $\mu$m binned down as expected for Gaussian noise. Wavelengths longer than this demonstrated correlated noise beyond $\sim$15 minutes. We applied the same fitting routine to the \texttt{ExoTic-JEDI} reduced spectroscopic light curves for the final transmission spectrum comparison. 

\subsubsection{Limb Darkening Impact}

\begin{figure*}
    \centering
    \includegraphics[width=0.9\linewidth]{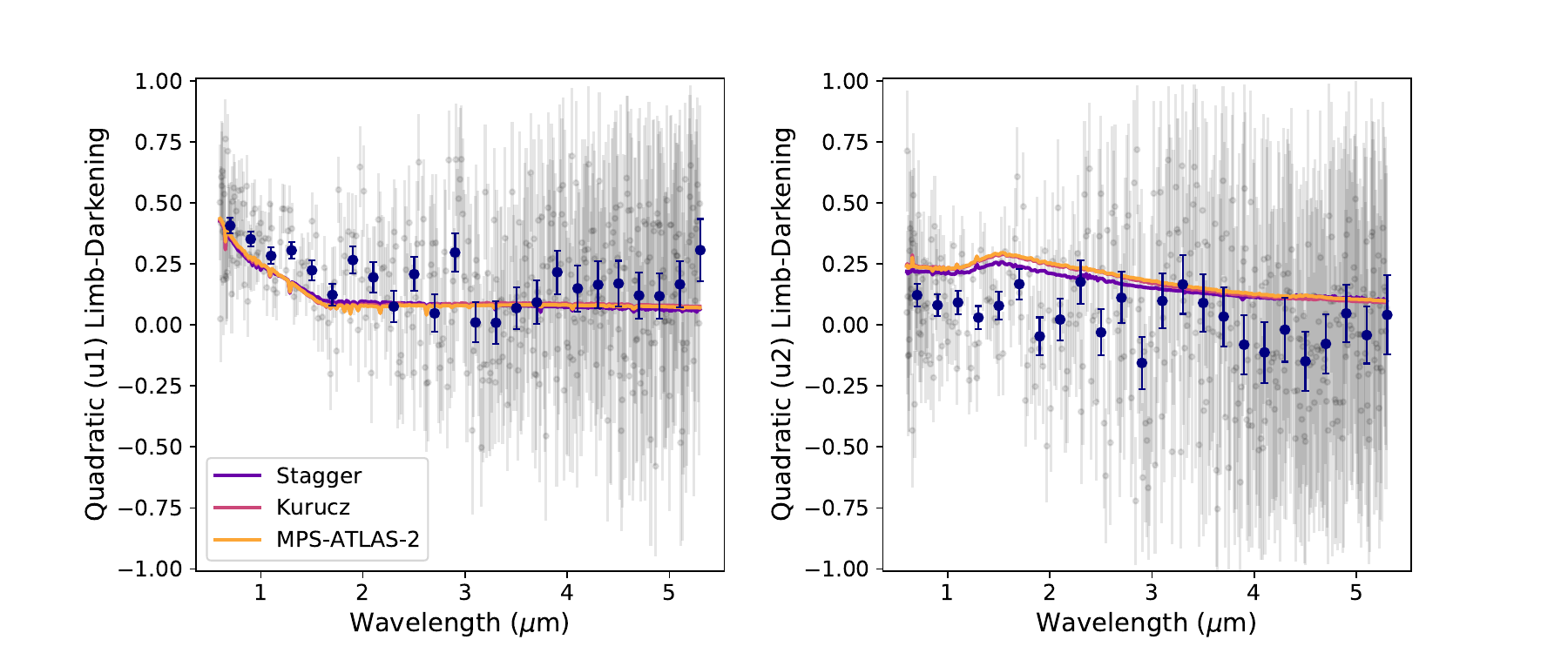}
    \caption{The best-fit quadratic limb-darkening parameters \citep[not following the re-parameterization detailed in][]{kipping2013} for the individual channels (\textit{black} and binned to 0.2 $\mu$m for clarity (\textit{blue}). The expected limb darkening parameters based on \citet{magic2015stagger} (\textit{purple}), \citet{kurucz1993atlas9} (\textit{red}), and \citet{kostogryz2023mps} (\textit{orange}) models are also shown. While the overall shape of the best-fit limb darkening parameters is in agreement for the bluer $<$ 4.5 $\mu$m, within the errorbars, there is an offset and significant scatter in some of the channels.}
    \label{fig:limbdark}
\end{figure*}

None of the best-fit first, second, or third-order limb darkening parameters matched the expected model values derived from Kurucz \citep{kurucz1993atlas9}, Stagger \citep{magic2015stagger}, or MPS2-ATLAS \citep{kostogryz2023mps} generated by \texttt{Exotic-LD} \citep{exoticld} assuming the stellar parameters for Kepler-51 \citep{libbyroberts2020}. For instance, Figure~\ref{fig:limbdark} plots the best-fit quadratic limb darkening terms, u1 and u2, against the expected model grids.

\citet{coulombe.limbdark.jwst.bias} concluded this as a possibility as low SNR transits (such as the JWST individual spectroscopic channels) do not possess the precision required for a more informative limb-darkening law. We bin the spectra into 80 nm bin light curves (59 total light curves). We repeat the fit routine finding identical limb darkening values to those derived from the native fitting. Therefore, the SNR at native resolution is not causing the deviation from limb darkening models.

Most stellar models are calculated assuming an older, inactive star. As shown by \citet{magneticfields.limbdark}, magnetically active regions can impact the modeled limb darkening values. Kepler-51, as a young active star, may not be well defined by previous limb darkening models. Moreover, the small stellar spot on the ingress limb of the transit could also effect the measured limb darkening due to differing wavelength-dependent contrasts. Attempts to fix the wavelength limb-darkening parameters significantly impacted the overall transmission spectrum by introducing unusual and nonphysical `turnover' in transit depth for the bluest wavelengths. Regardless, the overall slope in the transmission spectrum remains robust. Given the potential uncertainties in the stellar models for this young G dwarf, we opted to fit the quadratic limb darkening parameters.

\subsection{Spectroscopic Light Curves - Starry}\label{sec:spectroscopiclc_starry}
We also performed spectroscopic light curve fitting with \texttt{chromatic\_fitting} and \texttt{starry}. At high orders of spherical harmonics \texttt{starry} is very computationally expensive. Therefore, we binned the spectroscopic light curves to a low resolution ($\lambda/\delta \lambda$=10), resulting in 26 light curves. Holding the white light curve parameters from Section \ref{sec:wlc_starry} fixed, we fit for only the wavelength-dependent parameters; the three polynomial parameters, a linear limb-darkening parameter, planet radius, spot contrast, and error inflation term. We used the same priors for the fitted parameters as in Section \ref{sec:wlc_starry}, except with a uniform prior from 0--1 for the limb-darkening term. We chose to adopt a linear limb-darkening for the reasons described in Section \ref{sec:spectroscopiclc}. For each of the 26 spectroscopic light curves we performed MCMC sampling as in Section \ref{sec:wlc_starry}, this time with 500 tuning steps and 2000 draws across 2 chains.

\section{Transmission Spectrum}\label{sec:transmission_spectrum}

\begin{figure*}
    \centering
    \includegraphics[width=0.8\linewidth]{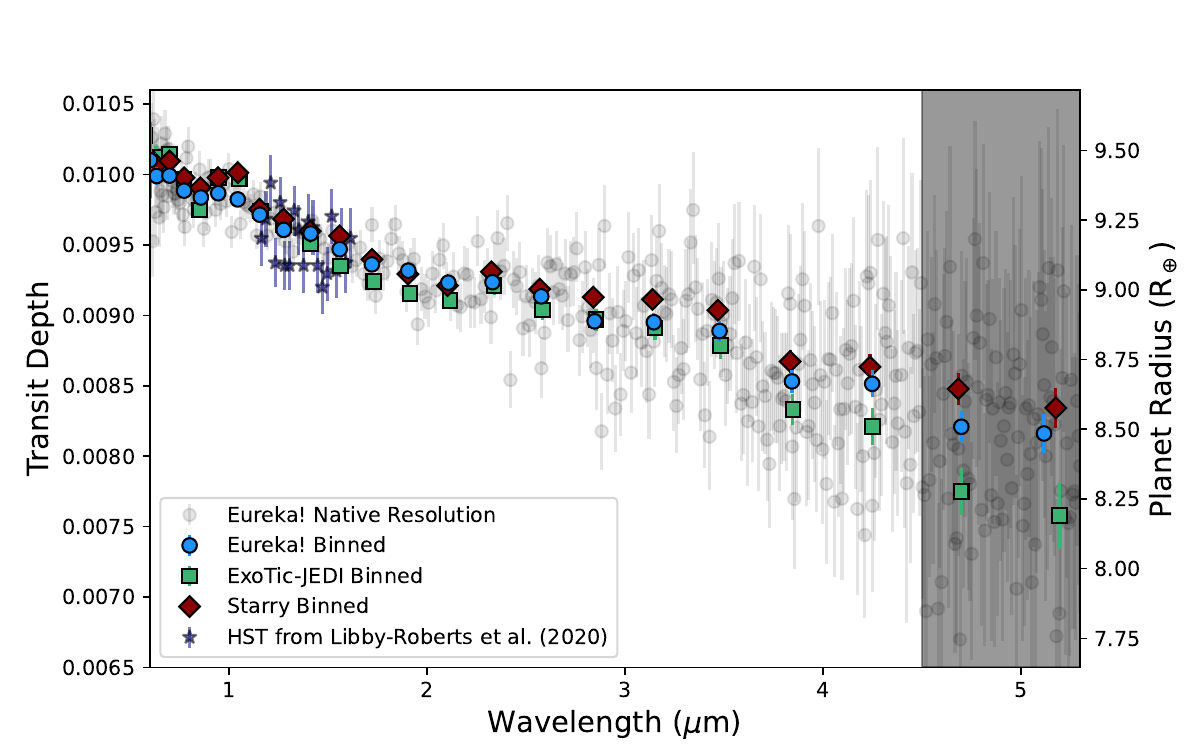}
    \caption{Kepler-51d's transmission spectrum observed with JWST/NIRSpec-PRISM covering 0.6 - 5.3 $\mu$m. The numerous gray points are the best-fit transit depths from the \texttt{Eureka!} reductions assuming a native resolution. These points were then binned to a R$\sim$10 plotted as blue circles. The same technique was applied to the \texttt{ExoTic-JEDI} fits (green squares). The \texttt{starry} spectrum (red diamonds) was derived by first binning the spectroscopic light curves to this resolution, and then fitting the data. For our analysis, we ignore wavelengths $>$ 4.5 $\mu$m as this spectroscopic light curves demonstrated significant noise producing large scatter on the transit depth fits. We've also included the HST/WFC3 transmission spectrum (light blue stars) presented in \citet{libbyroberts2020} for comparison. No offset was applied between the HST and JWST transit depths.}
    \label{fig:transmission_spectrum_comparison}
\end{figure*}

The final JWST NIRSpec-PRISM transmission spectrum derived from the \texttt{Eureka!}, \texttt{ExoTic-JEDI}, and \texttt{chromatic\_fitting} reductions and fits is plotted in Figure~\ref{fig:transmission_spectrum_comparison} with best-fit parameters listed in Table~\ref{tab:spectra} for the native \texttt{Eureka!} reduction and \texttt{spotrod} fits. We found that all pipelines and fitting routines led to a comparable spectrum - a near-featureless slope between 0.6--5.3 $\mu$m, from a transit depth of 1\% at the bluer wavelengths (planet radius of $\sim$9.25 R$_\oplus$) to a transit depth of 0.8\% at the redder wavelengths (planet radius of $\sim$8.5 R$_\oplus$). We also included the HST/WFC3 points from \citet{libbyroberts2020} in Figure~\ref{fig:transmission_spectrum_comparison} for comparison. We ignored points beyond 4.5 $\mu$m from further analysis as the scatter significantly increased within the individual light curves, and we begin to observe transit depth deviations between \texttt{chromatic\_fitting}, \texttt{Eureka!}, and \texttt{ExoTic-JEDI}.

\begin{table*}[]

\begin{tabular}{ccccc}
Wavelength ($\mu$m) & Transit Depth (\%)       & Limb Darkening (u$_1$)       & Limb Darkening (u$_2$)        & Spot Contrast                \\ \hline
0.6014              & 1.01$\pm$ 0.03  & 0.50$^{+0.24}_{-0.23}$ & 0.05$^{+0.33}_{-0.34}$  & 0.84 $\pm$ 0.05 \\
0.6056              & 1.00 $\pm$ 0.03 & 0.00$^{+0.15}_{-0.21}$ & 0.71$^{+0.30}_{-0.20}$  & 0.84 $\pm$ 0.05 \\
0.6100              & 1.02$\pm$ 0.03  & 0.42$^{+0.23}_{-0.23}$ & 0.12$^{+0.34}_{-0.33}$  & 0.75 $\pm$ 0.05 \\
0.6146              & 0.95$\pm$ 0.03  & 0.61$^{+0.25}_{-0.21}$ & -0.13$^{+0.31}_{-0.35}$ & 0.89 $\pm$ 0.05 \\
0.6192              & 1.03$\pm$ 0.02  & 0.57$^{+0.23}_{-0.22}$ & -0.12$^{+0.30}_{-0.33}$ & 0.86 $\pm$ 0.05 \\
0.6239              & 1.01$\pm$ 0.02  & 0.63$^{+0.21}_{-0.10}$ & -0.06$^{+0.28}_{-0.31}$ & 0.72 $\pm$ 0.04 \\
0.6287              & 0.97$\pm$ 0.02  & 0.38$^{+0.23}_{-0.21}$ & 0.30$^{+0.32}_{-0.33}$  & 0.82$^{+0.04}_{-0.05}$
\end{tabular}
\caption{Sample of the best-fit wavelength-dependent parameters from \texttt{Eureka!} with the native binning - all 408 channels are provided in machine-readable format. The spot contrast is assumed to be 0 for dark spot and 1 for spot temperature akin to photosphere (equal contrast).}\label{tab:spectra}
\end{table*}

The 0.75 R$\oplus$ planetary radii change across the spectrum corresponds to a change of three atmospheric scale heights. The overall slope and near featureless transmission spectrum of Kepler-51d are currently unique compared to other gas giant spectra \citep[e.g.][]{rustamkulov2022,bell.wasp80.jwst,sing.wasp107}. We conducted atmospheric retrievals and generated a grid of forward models to explore the origins of this super-puff's unusual spectrum.

\subsection{ExoTR Retrievals}\label{sec:exotr}

We used \texttt{ExoTR}, a fully Bayesian inverse retrieval framework (Tokadjian et al. 2025, in prep), to interpret the native resolution transmission spectrum in the range 0.6 to 4.5 $\mu$m. In \texttt{ExoTR}, the atmospheric structure is modeled through molecular and haze opacities, cloud contribution, Rayleigh scattering, collision induced absorption (CIA), and stellar contamination (e.g, \citealt{damiano.lhs1140b,bello-arufe.l98}). This is coupled with the nested sampling tool, MultiNest \citep{feroz2009}, to statistically infer the best fitting model to the data. In this work, we use 1000 live points with an error tolerance of 0.1. The planet radius and molecular abundances are drawn from a uniform prior: $R_p$ $\in$ $\mathcal{U}(0.5, 2) \times R_p$ where $R_p=0.844$~$R_J$ and CLR$_X$ $\in \mathcal{U}(-25, 25)$ where CLR is the centered log ratio which is converted to volume mixing ratio (VMR) space (see \citealt{damiano2021}). The planet atmosphere is assumed isothermal with $T_p=367$~K and the background fill-in gas is an 80\%-20\% mix of H$_2$-He. Also, the planet mass is a constant $M_p = 5.7$~$M_\oplus$. Stellar activity in \texttt{ExoTR} is modeled by calculating stellar contamination spectra using Phoenix models \citep{husser.phoenix}. We use 3 free parameters with the following prior distribution: the spot fraction $\delta_{\text{spot}} \in \mathcal{U}(0,0.5)$, faculae fraction $\delta_{\text{fac}} \in \mathcal{U}(0,0.5)$, spot temperature $T_{\text{spot}} \in \mathcal{U}(1500,T_{\text{phot}})$~K, faculae temperature $T_{\text{spot}} \in \mathcal{U}(T_{\text{phot}},1.2\times T_{\text{eff}})$~K, and photosphere temperature $T_{\text{phot}} \in \mathcal{N}$(5670, 60) (normal distribution with mean 5670~K and variance 60~K). For hazes, we consider both organic tholins \citep{khare1984} and soot \citep{OpticalPropertiesofAerosolsandCloudsTheSoftwarePackageOPAC} and parameterize the opacity by number density, which is translated into a relative volume mixing ratio, and particle size, where we assume the particles to be spherical. The mixing ratio and particle diameter follow a vertically uniform profile and are free parameters of the retrievals, with priors uniform in log-space VMR$_{\text{haze}} \in \mathcal{LU}(-10, -1)$ and $d_{\text{haze}} \in \mathcal{LU}(-3, 2)$ $\mu$m, respectively. We start with a base case of either a featureless flat spectrum or only stellar features and build up complexity by adding hazes and molecules while noting the increase in Bayesian evidence with each step.

\begin{figure}
    \centering
    \includegraphics[width=.8\linewidth]{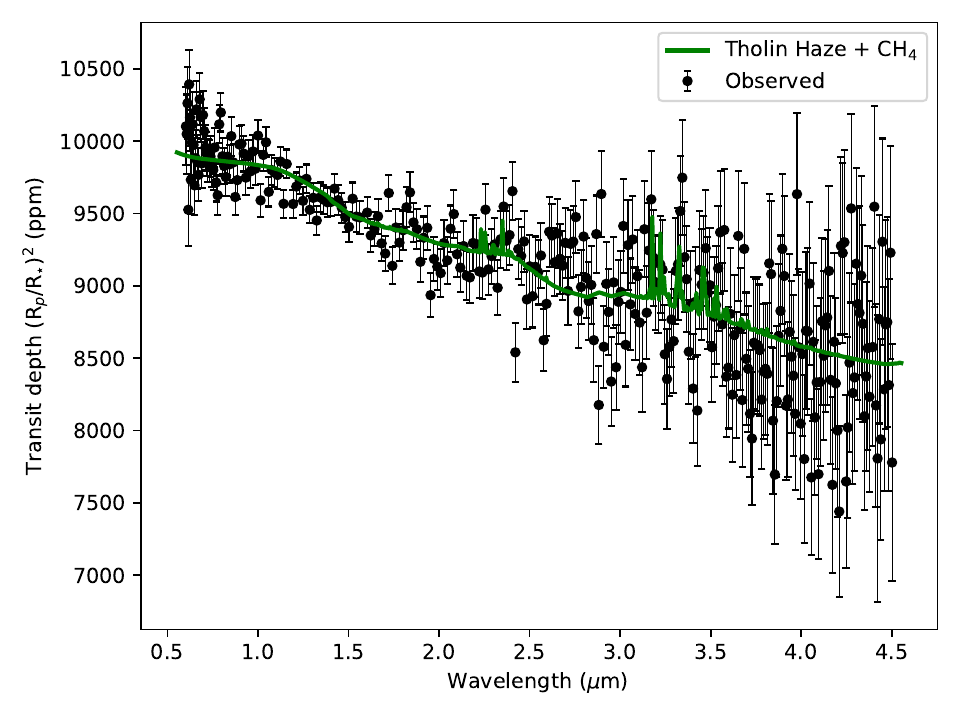}
    \caption{The best fit model from \texttt{ExoTR} plotted against the spectrum of Kepler-51d. This model is the tholin haze plus CH$_4$ case as highlighted in Table~\ref{tab:exotrresult}.}
    \label{fig:bestfit_exotr}
\end{figure}

Starting with a flat spectrum as the base case, we find that including either a tholin or soot haze improves the fit by $>$ 5$\sigma$. Adding H$_2$O, CO$_2$, HCN, and NH$_3$ one at a time in addition to haze does not increase the evidence. However, a tholin haze plus CH$_4$ case is preferred over the haze-only case by 2.2$\sigma$.(Table~\ref{tab:exotrresult}). The best fit model incorporates a tholin haze particle diameter of $0.52 \pm 0.04$ $\mu$m with an atmospheric abundance (volume mixing ratio) of $10.2\pm 2.6$ ppb (mass mixing ratio of $7.09 \times 10^{-7}$) and a CH$_4$ mixing ratio of 66$^{+72}_{-45}$ ppb. The best fit model from \texttt{ExoTR} is shown overlain on the data in Figure~\ref{fig:bestfit_exotr} and pertains to the tholin haze + CH$_4$ scenario.

If we assume possible stellar heterogeneity spectrum as the base case (fitting for both the photosphere temperature, spot temperature and coverage fraction), we find again that a hazy atmosphere improves the evidence by $>$ 5$\sigma$. However, adding any molecule as a free parameter, even CH$_4$, does not lead to any gain in evidence. Thus, although there is a hint of CH$_4$ detection in the case above, there is not enough evidence to confirm its presence in the atmosphere. In this case, the haze particles have diameter $0.67 \pm 0.08$ $\mu$m and fraction 32 $^{+160}_{-12}$ ppb for the best fit. We also find an unrealistic cold spot temperature of 3100 K, implying a flux contrast in excess of 0.9 using the \texttt{starry} method for this G Dwarf. We therefore are confident that the transmission spectrum is unaffected by significant stellar heterogeneities (see Section~\ref{sec:ootstellarspectrum}). 

\begin{table}[h!]
\centering
\begin{tabular}{lcl}
\toprule
\textbf{Scenario} & \textbf{Log Evidence} & \textbf{$\sigma$ baseline} \\
\hline
Tholin Haze + CH\textsubscript{4}    & 2058.06 & 2.2 \\
Tholin Haze + HCN                    & 2056.35 & -- \\
Tholin Haze + H\textsubscript{2}O    & 2056.28 & -- \\
Tholin Haze + CO\textsubscript{2}    & 2056.18 & -- \\
Tholin Haze + NH\textsubscript{3}    & 2055.60  & -- \\
Tholin Haze                          & 2056.85 & Base case \\
\hline
Tholin Haze                          & 2056.85 & $>$ 5 \\
Soot Haze                            & 2056.69 & $>$ 5 \\
Flat Spectrum                        & 1536.52 & Base case \\
\end{tabular}
\caption{Summary of retrieval results with \texttt{ExoTR} }\label{tab:exotrresult}
\end{table}

\subsection{PLATON Retrievals}

We use \texttt{PLATON} \citep{platon1,platon2} to perform retrievals on the native resolution spectrum of Kepler-51d. The posteriors and evidence are calculated with the nested sampling technique using \texttt{dynesty} \citep{dynesty1}. Given that the native resolution of the spectrum can reach R $\sim 500$, we use R=10,000 opacities for our retrievals since \texttt{PLATON} uses opacity sampling to calculate the transmission spectrum. The planet mass and stellar radius are fixed to their best fit updated values in \citet{masuda.kepler51e} ($5.6 \, M_\oplus$ and $0.881 \, R_\odot$, respectively). In our fiducial retrieval, we fit for the planet radius at 1 mbar (with prior $\mathcal{U}(0.6, 1) \times R_{\rm p}$, with $R_{\rm p} = 0.844~R_{\rm J}$), the atmospheric temperature in K (assumed isothermal, $\mathcal{U}(100, 400)$), log of the metallicity in terms of solar metallicity ($\mathcal{U}(-1, 3)$), and haze properties: log of the median particle size in meters ($\mathcal{U}(-8, -6)$), vertically constant mixing ratio (parameterized as number density in m$^{-3}$ at 1 kbar with prior $\mathcal{U}(10, 25)$), and log of the imaginary refractive index $k$ ($\mathcal{U}(-3, 1)$, assumed to be wavelength independent). The particle size distribution is assumed to be log-normal with a width of 0.5. The real refractive index is fixed to 1.65 to roughly match the corresponding value for soots and tholins in the relevant wavelength range \citep[e.g.][]{titan.haze.optical.prop}. We do not fit for the carbon-to-oxygen ratio (C/O) of the atmosphere as no molecular species are retrieved with a confidence $>$ 3$\sigma$. We also perform retrievals in which we fix the refractive indices to those of soots \citep{morley2015}, tholins \citep{khare1984}, and sulfur hazes (S$_8$, \citealt{fuller1998}, $k$ is negligibly small ($< 10^{-6}$) in the $2-6~\mu$m range so its value is obtained from a spline interpolation) and let the other particle properties (number density, particle size) vary. This allows us to have wavelength-dependent refractive indices and to simulate the effect of this chromatic dependence by varying haze physical properties.

\begin{figure}
    \centering
    \includegraphics[width=.8\linewidth]{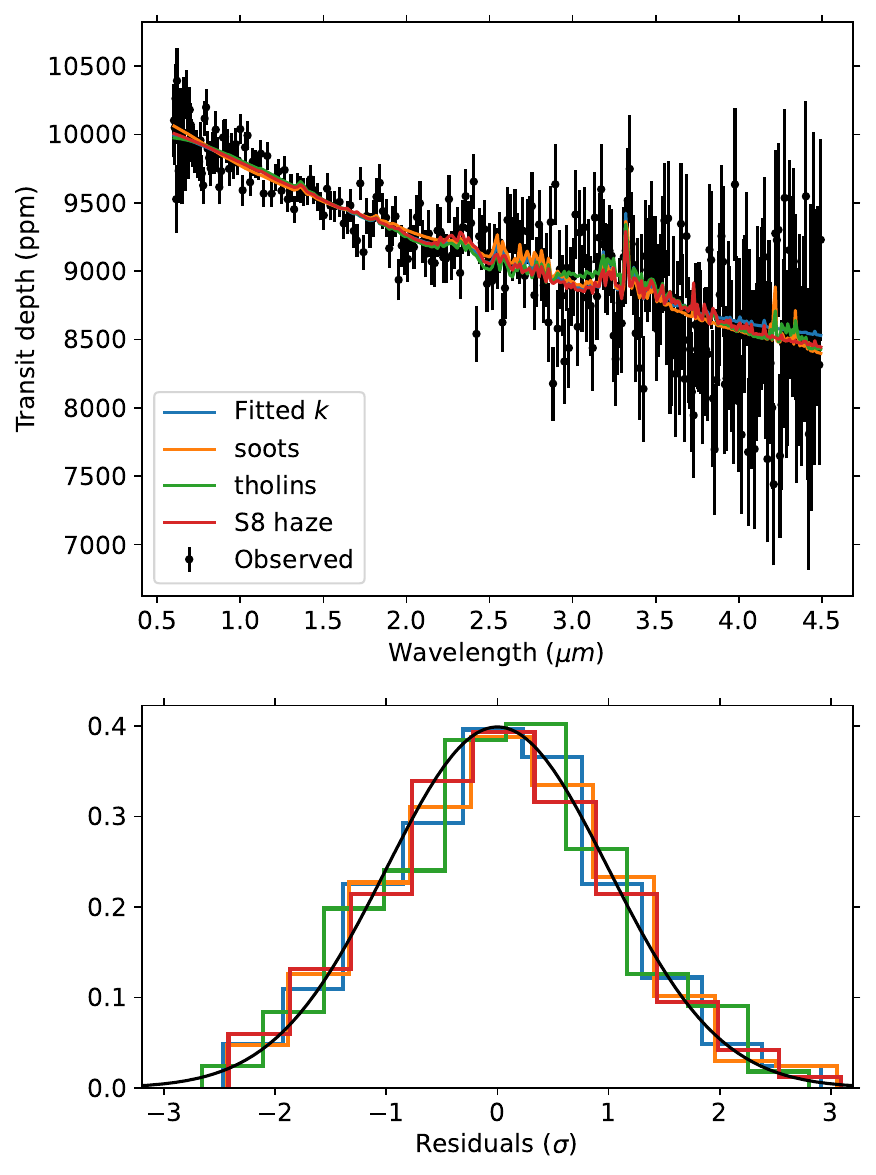}
    \caption{The best fit models for the transmission spectrum of Kepler-51 d. The fiducial retrieval fits a wavelength-independent imaginary refractive index $k$ and the other retrievals assume refractive indices for a specific haze. The bottom panel shows the residuals for each best fit model.}
    \label{fig:platon_spectrum_fits}
\end{figure}

Figure~\ref{fig:platon_spectrum_fits} shows the best fit models for our different retrievals along with a histogram of the residuals. The histogram of the residuals follows the expectation from normally distributed errors without the need for error inflation. The evidence for retrievals with wavelength-independent $k$, sulfur hazes, and soots is comparable and the data are unable to distinguish between these scenarios. Only tholins are rejected by the data with a Bayes factor of $\sim 14$, primarily because of additional absorption features in the $2-4 ~\mu$m region that are not supported by the data. We also perform a retrieval in which we fit a wavelength-independent $k$ but turn off gas absorption to quantify the significance of molecular detection in the spectrum: this yields a Bayes factor of $\sim 1$, implying that the absence of molecular absorption leads to an equivalently good fit to the data. The haze properties retrieved without gas absorption are consistent with the results of the fiducial retrieval.

\begin{figure}
    \centering
    \includegraphics[width=.8\linewidth]{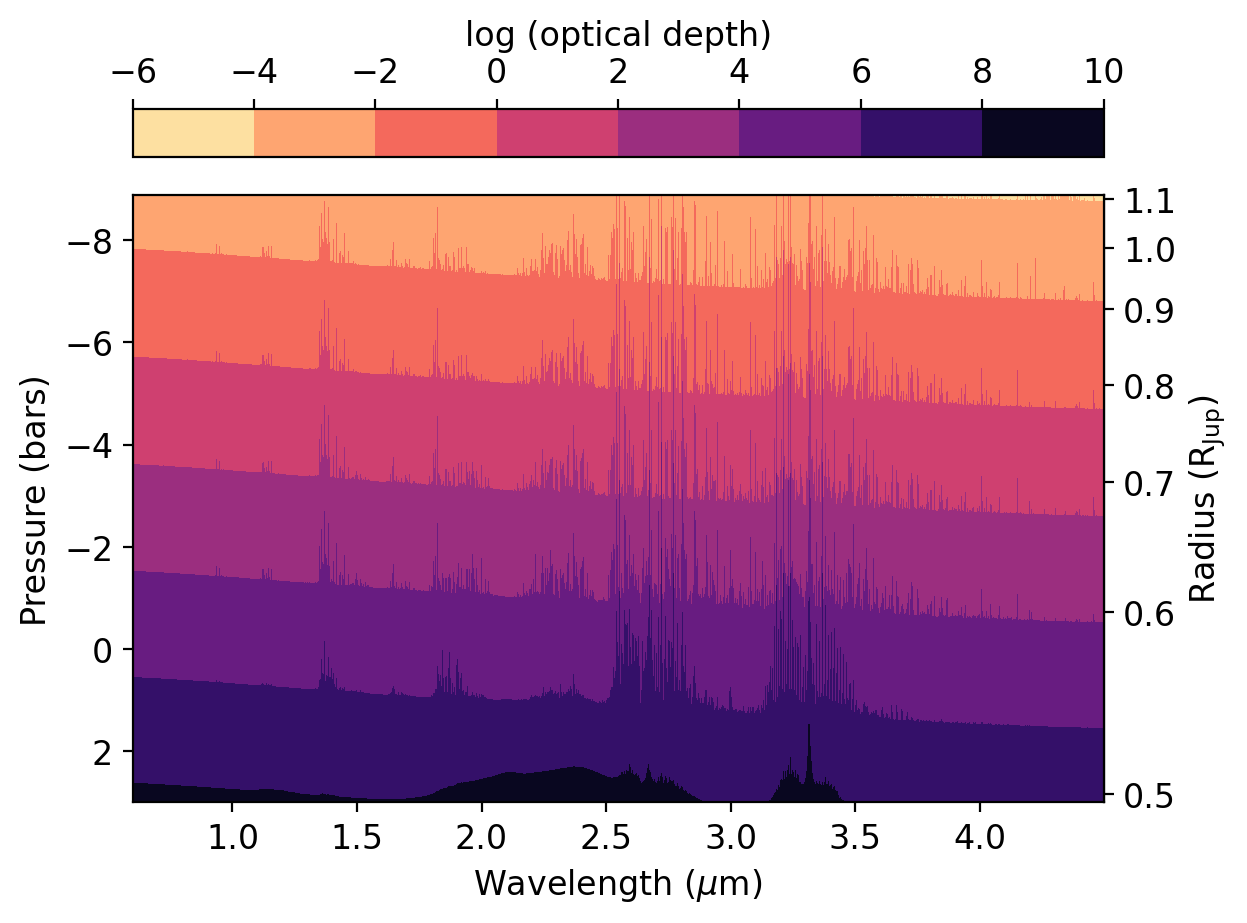}
    \caption{Slant optical depth $\tau$ (color) as a function of wavelength and both the pressure (left) and radius (right) in the best-fit model in the fiducial retrieval. The $\tau \sim 1$ surface corresponds to a pressure range of $1-10 \, \mu$ bar.}
    \label{fig:tau_pressure_radius}
\end{figure}

All our retrievals suggest that the pressure corresponding to the observed radius of the planet is significantly lower than the typically assumed range of $1 - 100$ mbar in transit geometry. Figure~\ref{fig:tau_pressure_radius} shows the optical depth as a function of wavelength, pressure, and planet radius for the best fit model in our fiducial retrieval. The haze makes the atmosphere optically thick ($\tau = 1$) at a pressure of $\sim 1-10 \, \mu$bar. The optical radius of $0.844 \, R_{\rm Jup}$ corresponds to a pressure of $\sim 1 \, \mu$bar. Models with higher photospheric pressures are rejected by \texttt{PLATON} because the planet radius at 1 nbar becomes larger than the Hill radius of the planet. 

\subsection{Forward Modeling} \label{sec:forward_haze}

We simulated the radius and density distributions of the haze particles, as well as the corresponding transmission spectra, using the haze microphysical and transmission spectrum models of \citet{2018ApJ...853....7K}.
We computed model grids by varying atmospheric metallicity, haze monomer production rate, eddy diffusion coefficient, and haze optical properties.
For atmospheric metallicity, we considered three cases--1, 10, and 100$\times$ the solar value--adopting solar elemental abundances from \citet{2003ApJ...591.1220L}.
The haze monomer production rate, $\dot{M}$, was explored in 1-dex increments, ranging from $10^{-15}$ to $10^{-10}$~$\mathrm{g}$~$\mathrm{cm}^{-2}$~$\mathrm{s}^{-1}$, along with a haze-free (clear) atmosphere case.
The monomer production profile was modeled using a log-normal distribution, following Eq.~(10) of \citet{ohno2020} with the same mean and standard deviation values.
Vigorous vertical mixing, strong enough to dominate over sedimentation transport, leads to a steeper spectral slope \citep{kawashima2019b, ohno2020}. Therefore, we considered two values for the eddy diffusion coefficient: $K_\mathrm{zz} = 10^7$ and $10^9$~$\mathrm{cm}^2$~$\mathrm{s}^{-1}$.
Finally, given the uncertainty in the refractive index of haze in exoplanetary atmospheres, we examined two representative cases, the same as those used in the spectral retrieval with \texttt{ExoTR} (see Section~\ref{sec:exotr}): tholin \citep{khare1984} and soot \citep{ OpticalPropertiesofAerosolsandCloudsTheSoftwarePackageOPAC}.

The haze monomer radius ($s_1$) and internal density ($\rho_p$) were set to $s_1 = 1$~nm and $\rho_p = 1.0$~g~$\mathrm{cm}^{-3}$, respectively.
For the temperature-pressure profile, we adopted the publicly available\footnote{\url{https://cdsarc.u-strasbg.fr/viz-bin/qcat?J/A+A/562/A133}} analytical model of \citet{Parmentier2014}, assuming an internal temperature of 100~K and applying their correction factor of 0.25 (planet-average profile), with the default coefficient and opacity settings \citep{Parmentier2015, Valencia2013}.
We used the temperature-pressure profile for a solar-metallicity atmosphere for both the 10 and 100 $\times$ solar metallicity cases due to the lack of available opacity options.
Gaseous abundance profiles were calculated using the photochemical model of \citet{2018ApJ...853....7K}, adopting the solar UV spectrum provided by the VPL team \citep{2003AsBio...3..689S} \footnote{\url{https://vpl.astro.washington.edu/spectra/stellar/other_stars.htm}}, given that the spectral type of Kepler-51 \citep[G4 E;][]{2022ApJS..261...26S} is similar to the Sun.
For simplicity, the gaseous abundance profiles calculated with $K_\mathrm{zz} = 10^7$~$\mathrm{cm}^2$~$\mathrm{s}^{-1}$ were also used for the haze microphysical calculations with $K_\mathrm{zz} = 10^9$~$\mathrm{cm}^2$~$\mathrm{s}^{-1}$, as the dependence of the gaseous abundance profiles on $K_\mathrm{zz}$ is relatively weak compared to that of haze profiles.

\begin{figure}
    \centering
    \includegraphics[width=.8\linewidth]{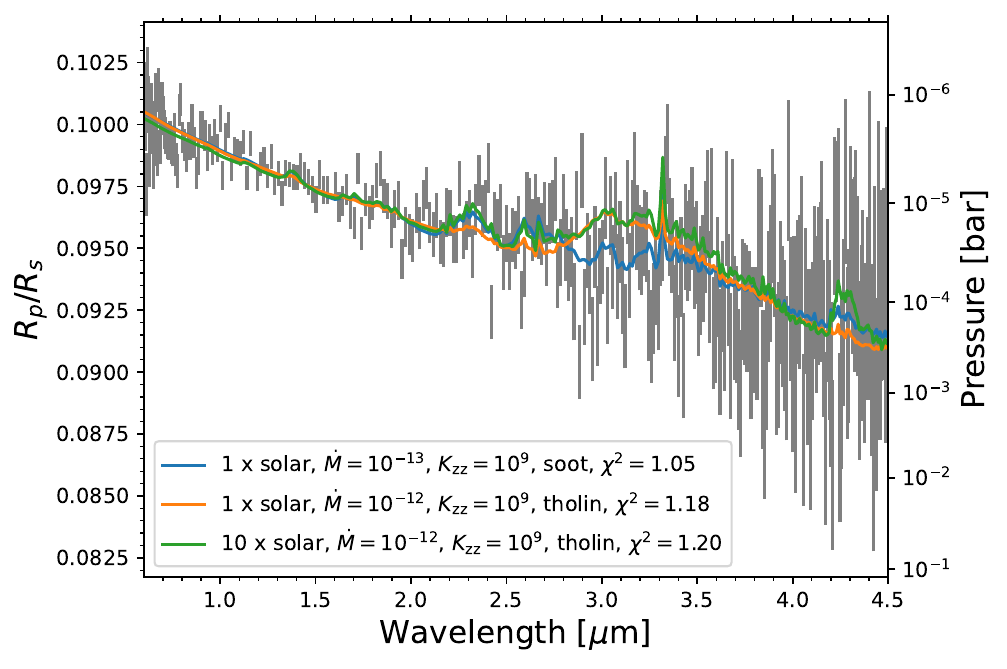}
    \caption{Spectrum models for the three best-fit cases. The right vertical axis indicates the corresponding pressure levels for the best-fit model (blue line), which assumes a metallicity of 1$\times$ solar, a monomer production rate of $\dot{M} = 10^{-13}$~$\mathrm{g}$~$\mathrm{cm}^{-2}$~$\mathrm{s}^{-1}$, an eddy diffusion coefficient of $K_\mathrm{zz} = 10^9$~$\mathrm{cm}^2$~$\mathrm{s}^{-1}$, and soot optical properties.}
    \label{fig:forward_haze_spectrum}
\end{figure}

\begin{figure*}
\gridline{\fig{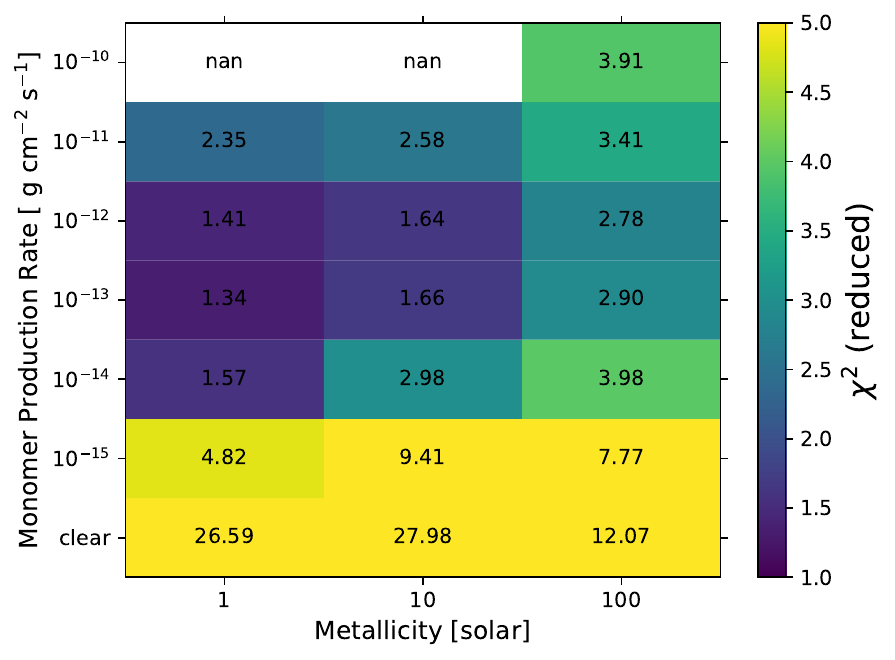}{0.45\textwidth}{(a) $K_\mathrm{zz} = 10^7$ $\mathrm{cm}^2$ $\mathrm{s}^{-1}$, tholin}
          \fig{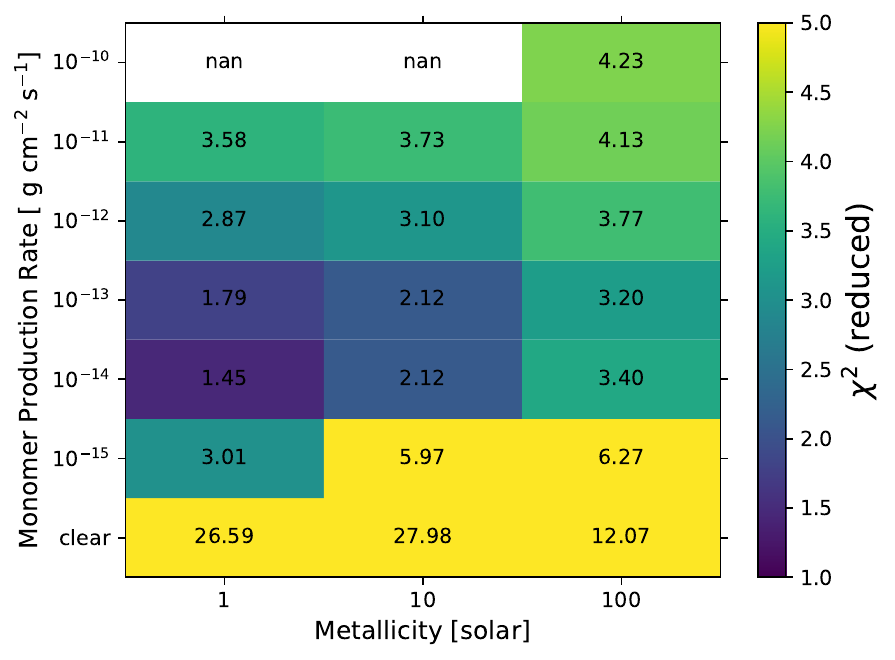}{0.45\textwidth}{(b) $K_\mathrm{zz} = 10^7$ $\mathrm{cm}^2$ $\mathrm{s}^{-1}$, soot}
          }
\gridline{\fig{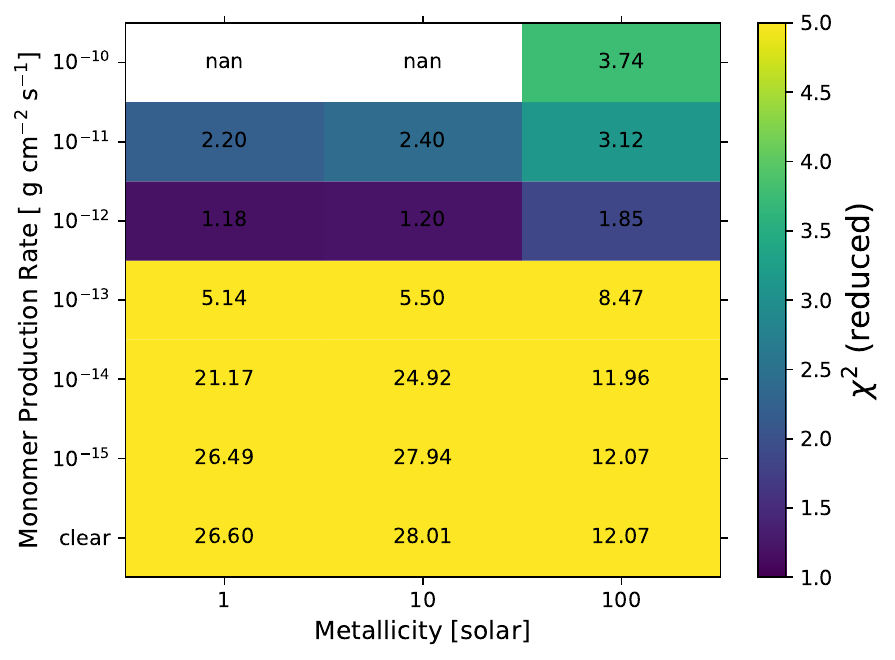}{0.45\textwidth}{(c) $K_\mathrm{zz} = 10^9$ $\mathrm{cm}^2$ $\mathrm{s}^{-1}$, tholin}
          \fig{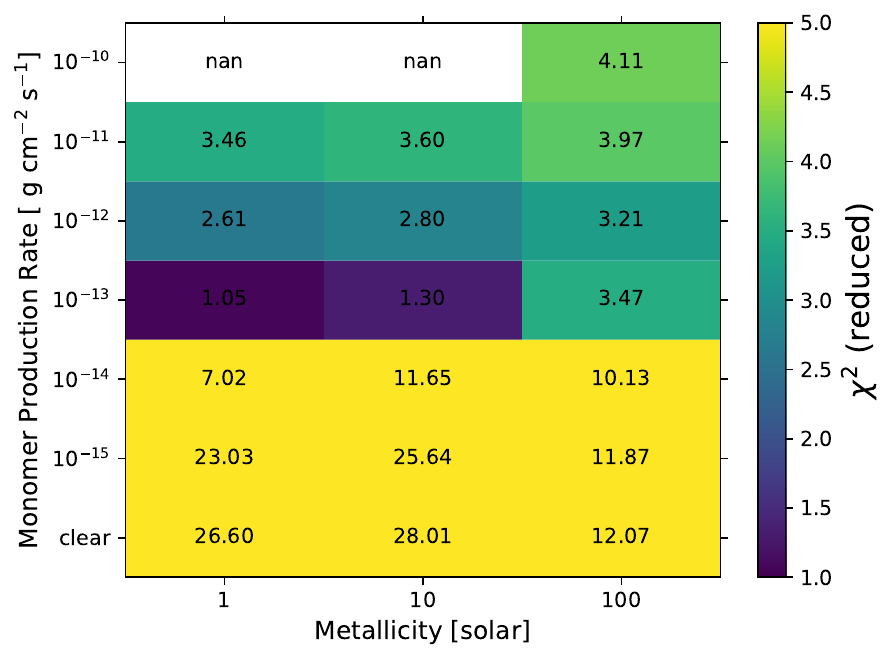}{0.45\textwidth}{(d) $K_\mathrm{zz} = 10^9$ $\mathrm{cm}^2$ $\mathrm{s}^{-1}$, soot}
          }
\caption{Goodness of fit for the parameter space (atmospheric metallicity, haze monomer production rate, eddy diffusion coefficient, and haze optical properties) explored using the haze models. Note that the haze microphysical simulation failed to converge for cases with a haze monomer production rate of $10^{-10}$~$\mathrm{g}$~$\mathrm{cm}^{-2}$~$\mathrm{s}^{-1}$ in the 1$\times$ and 10$\times$ solar metallicity cases due to the extremely high monomer production rate.
\label{fig:forward_haze_chi2}}
\end{figure*}

The spectrum models for the three best-fit cases are presented in Figure~\ref{fig:forward_haze_spectrum}.
Additionally, the reduced $\chi^2$ values for the entire explored parameter space--namely, atmospheric metallicity, haze monomer production rate, eddy diffusion coefficient, and haze optical properties--are shown in Figure~\ref{fig:forward_haze_chi2}.
In calculating the $\chi^2$ values, we included only the data points at wavelengths below 4.5~$\mu$m, as noted at the beginning of this section.
It is also worth mentioning that the haze microphysical simulation failed to converge for cases with a haze monomer production rate of $10^{-10}$~$\mathrm{g}$~$\mathrm{cm}^{-2}$~$\mathrm{s}^{-1}$ in the 1$\times$ and 10$\times$ solar metallicity cases due to the extremely high monomer production rate.

As shown in Figure~\ref{fig:forward_haze_spectrum}, in the best-fit models, the spectral slope produced by haze particles successfully reproduces the observed spectral gradient.
The observed steep spectral gradient favors scenarios with low metallicity, a moderate monomer production rate, and a large eddy diffusion coefficient, as only this combination can account for its characteristics, as described below.

First, reproducing the observed large variation in transit depth requires low metallicity, which increases the atmospheric scale height.
Regarding the monomer production rate, a moderate rate yields the steepest spectral slope \citep{ohno2020}. This is because a high monomer production rate results in a wavelength-independent flat spectrum, whereas a low monomer production rate leads to a spectrum with prominent molecular features and a less steep spectral slope \citep{2017ApJ...847...32L, kawashima2019b}, both of which are inconsistent with the observed spectrum.
Additionally, a sufficiently large eddy diffusion coefficient is necessary for vertical transport by eddy diffusion to dominate over sedimentation, thereby producing a steep spectral slope \citep{ohno2020}.

Furthermore, among the simulated grids, the soot optical properties yielded the lowest $\chi^2$ value due to the apparent absence of the distinctive C-H bond absorption feature of tholin at $\sim3.0$~$\mu$m.

\subsection{Comparison Between Retrievals and Forward Models}

\texttt{PLATON} and \texttt{ExoTR} retrieve similar haze properties based on the native resolution transmission spectrum for Kepler-51d. Both models use an isothermal temperature profile, which we fit for. We determine a best-fit temperature slightly cooler than equilibrium temperature between 280 - 300 K compared to 350 K with zero albedo. This is unsurprising given the high altitude we are probing. However, if we assume this temperature to be the `true' equilibrium temperature, Kepler-51d requires an albedo between 0.3--0.6 --- similar to Titan\citep{titan.albedo} and Jupiter\citep{jupiteralbedo} -- both which have hazes in their atmospheres. However, both moon and planet do not have featureless spectra indicating Kepler-51d's haze layer must be substantially thicker. Kepler-51d's albedo value is also comparable to GJ 1214b's 0.51 +/- 0.06 which also displays a featureless albeit flat spectrum \citep{gj1214.jwstspectrum}. The retrievals are also in agreement with the low pressures probed in the transmission spectrum (1-10 $\mu$bars). 

The two retrievals differ in the methane abundance with \texttt{ExoTR} tentatively detecting methane while \texttt{PLATON} preferring no molecular features. Tholin hazes are disfavored by \texttt{PLATON} but remain viable under with \texttt{ExoTR}. However, both findings show only slight statistical deviations and should be interpreted with caution. As presented in the previous section, we also performed forward spectral modeling based on the haze microphysical model to compare with the retrieved results. Comparing the retrievals to the physically-motivated forward models, we see that all results prefer a 1-10$\times$ solar atmospheric metallicity set solely by the required scale height of the atmosphere to span the large planetary radius in the spectrum. Soot and tholin compositions lead to nearly the same $\chi^2_r$, although tholins are slightly less favored due to the apparent absence of their characteristic C-H bond absorption feature at $\sim3.0$~$\mu$m, consistent with the \texttt{PLATON} results.

\begin{figure*}
    \centering
    \includegraphics[width=0.49\linewidth]{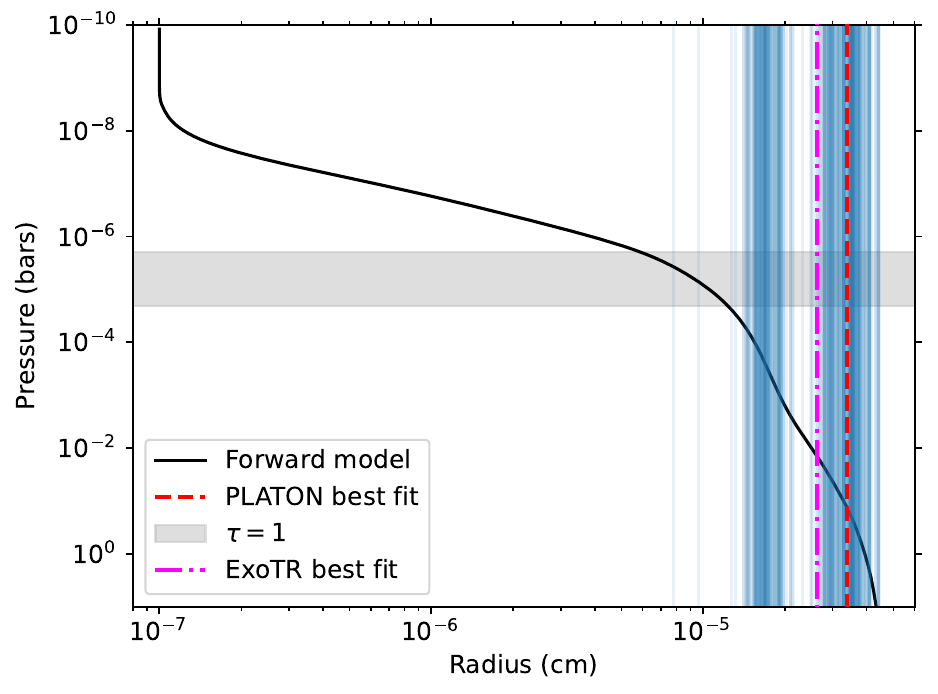}
    \includegraphics[width=0.49\linewidth]{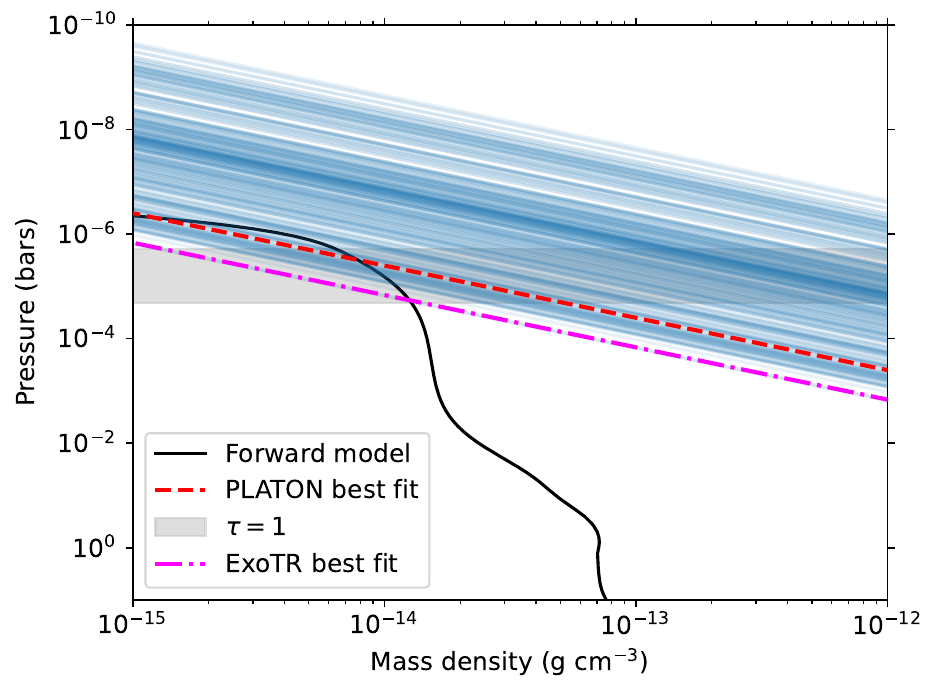}
    \caption{The volume averaged particle size and total mass density from the forward models compared to the corresponding values from retrieved models from \texttt{PLATON} and \texttt{ExoTR} with wavelength-independent $k$. The modeled and retrieved estimates agree well in the $\sim 1-10~\mu$bar region, where the retrieved transit photosphere is located.}
    \label{fig:forward_retrieval_comparison}
\end{figure*}

Both the retrievals and forward models indicate that Kepler-51d's spectrum probes unusually low pressures $\sim$1 $\mu$bars at the shorter wavelengths. However, while the retrievals span a pressure range from 1-10 $\mu$bars, the forward models suggest a pressure level of 100 $\mu$bars at longer wavelengths. For the pressure range that corresponds to $\tau = 1$ for the retrievals, the retrieved mass-averaged particle size and the total mass density of the haze agree quite well with the forward model (Figure~\ref{fig:forward_retrieval_comparison}). The retrieved haze sizes are slightly larger (0.15 - 0.35 $\mu$m instead of 0.1 $\mu$m) and the haze mass density is slightly higher, likely due to the featureless nature of the observed spectrum. 

These discrepancies are also due to the limitations of both retrieval codes, which assume a single particle size and number fraction across all pressure levels. In addition, \texttt{ExoTR} does not assume a distribution in particle size within each atmospheric layer but takes a constant haze particle size with density 0.8 g cm$^{-3}$. These assumptions, while not physically motivated, are required due to the level of complexity and computational intensity required for the retrieved atmospheric models. Our forward model in turn, considers physically motivated haze physics. Forward modeling performs computationally-expensive simulations to derive physically-motivated parameters including particle size and number fractions as functions of pressure. Given the physical nature of the forward model, we are inclined to conclude that Kepler-51d possesses a thick haze layer spanning two orders of magnitude in pressure of small $\sim$ 0.1 $\mu$m particles. These particle sizes are similar to Titan's tholin haze particle sizes at similar pressure levels \citep{titan.detatched.haze}. We also emphasize the importance of considering both forward models and retrievals when interpreting the complex spectra of planets such as Kepler-51d.

We note that in the above analyses, all haze particle sizes are assumed to be spherical. It is possible that these particles could possess a monomer structure similar to photochemical hazes on Titan and Pluto \citep[e.g. fluffy aggregates,][]{adams2019,ohno2020agg} which would impact their scattering properties. However, as we are able to reproduce Kepler-51d's spectrum assuming spherical particles, we leave fluffy aggregates for potential future exploration. 

\section{Discussion - Atmosphere of Kepler-51d}\label{sec:planet_discussion}

\subsection{Lack of Molecular Features}

At 350 K, we expect to observe a rich assortment of molecular features (methane, water, carbon dioxide, and ammonia) assuming an aerosol-free chemical equilibrium atmosphere for Kepler-51d \citep[e.g.][]{fortney.2020} -- notably given its extreme scale height of $\sim$1700 km. Instead, the lack of any clear detectable features in an extended H/He-rich atmosphere between 0.6--5.3 $\mu$m is a first for JWST. However, carbon-, oxygen-, nitrogen-, and potentially sulfur-based molecules must exist in Kepler-51d's atmosphere to seed haze formation. Beyond an analysis of the haze layer itself, we are unable to recover a C/O ratio or a measured atmospheric metallicity driven by absorption features in the spectrum. Assuming the 66$^{+72}_{-45}$ ppb detection of methane from Section~\ref{sec:exotr}, we hypothesize that the methane molecules are uplifted from deeper in the atmosphere and the precursor to haze production -- not an overall representation of the true methane abundance of Kepler-51d. Within this framework, it would suggest haze composition of carbon-based hazes (tholin or soot), though given the $<$ 3$\sigma$ confidence in methane detection, we cannot exclude sulfur-based hazes from the spectrum.

\subsection{The Inflated Nature of Kepler-51d}

\subsubsection{A High-Altitude Haze Layer}
Based on both retrievals and forward models, the transmission spectrum of Kepler-51d can be best characterized by a high altitude haze layer at pressures 1-100 $\mu$bars of either carbon-dominated or sulfur-dominated haze particles. This pressure range is akin to Titan's high altitude haze layer, providing a physical example that it is possible to maintain hazes at these low pressures \citep{titan.detatched.haze}. However, we can test the \citet{gao.and.zhang} hypotheses of an inflated radius due to hazes. Using \texttt{PLATON}, we approximate the planetary radius at 20 mbars motivated by \citet{lopez.and.fortney} to be $\sim$7.1 R$_\oplus$. This is a 1.8 R$_\oplus$ difference from the measured optical planetary radius 9.3 R$_\oplus$. We approximate the H/He-mass fraction required assuming this smaller 20 mbar radius using the thermal evolution models from \citet{lopez.and.fortney}. Based on \citet{lopez.and.fortney}, 30\% of Kepler-51d's mass is expected to comprise its H/He envelope. While less than the previous approximation of $>$ 40\% from \citet{libbyroberts2020}, this is still extreme, notably as \citet{lopez.and.fortney} also approximate a core mass of 3.9 M$_\oplus$. Therefore, the high-altitude haze-layer alone cannot explain the overall inflated radius of Kepler-51d. 
Thus, if hazes are responsible for the sloped spectrum of Kepler-51d, then the planet must also maintain an abnormally large H/He atmosphere.

\subsubsection{Interior Heating}

Recent JWST observations of the super-puff WASP-107b depict a methane-depleted atmosphere which \citet{sing.wasp107} and \citet{Welbanks2024} to derive a substantially hotter intrinsic temperature than expected. While Kepler-51d's spectrum is featureless, we can compare these two low-density planets. 

We calculate the tidal luminosity from Eq. (1) in \citet{millholland.2020} assuming the updated 0.05 eccentricity based on the four planet model in \citet{masuda.kepler51e}. Assuming 10$^{5}$ as the tidal quality factor Q and a k2 value of 1.5, we calculate a tidal luminosity on the order of 10$^{18}$ ergs s$^{-1}$. In turn, we determine the received stellar power to be 10$^{26}$ ergs s$^{-1}$, a factor 10$^{9}$ times larger than the tidal heating. \citet{millholland.2020} notes that L$_\mathrm{tide}$/L$_\mathrm{irr}$ $>$ 10$^{-5}$ is required for tides to significantly inflate a planet. For Kepler-51d the overall tidal luminosity is negligible. We therefore disfavor the hypothesis of a heated interior (beyond the current 500 Myr age) for the mechanism behind Kepler-51d's inflated radius.

\subsubsection{Possibility of Circumplanetry Ring}

\begin{figure*}
    \centering
    \includegraphics[width=0.48\linewidth]{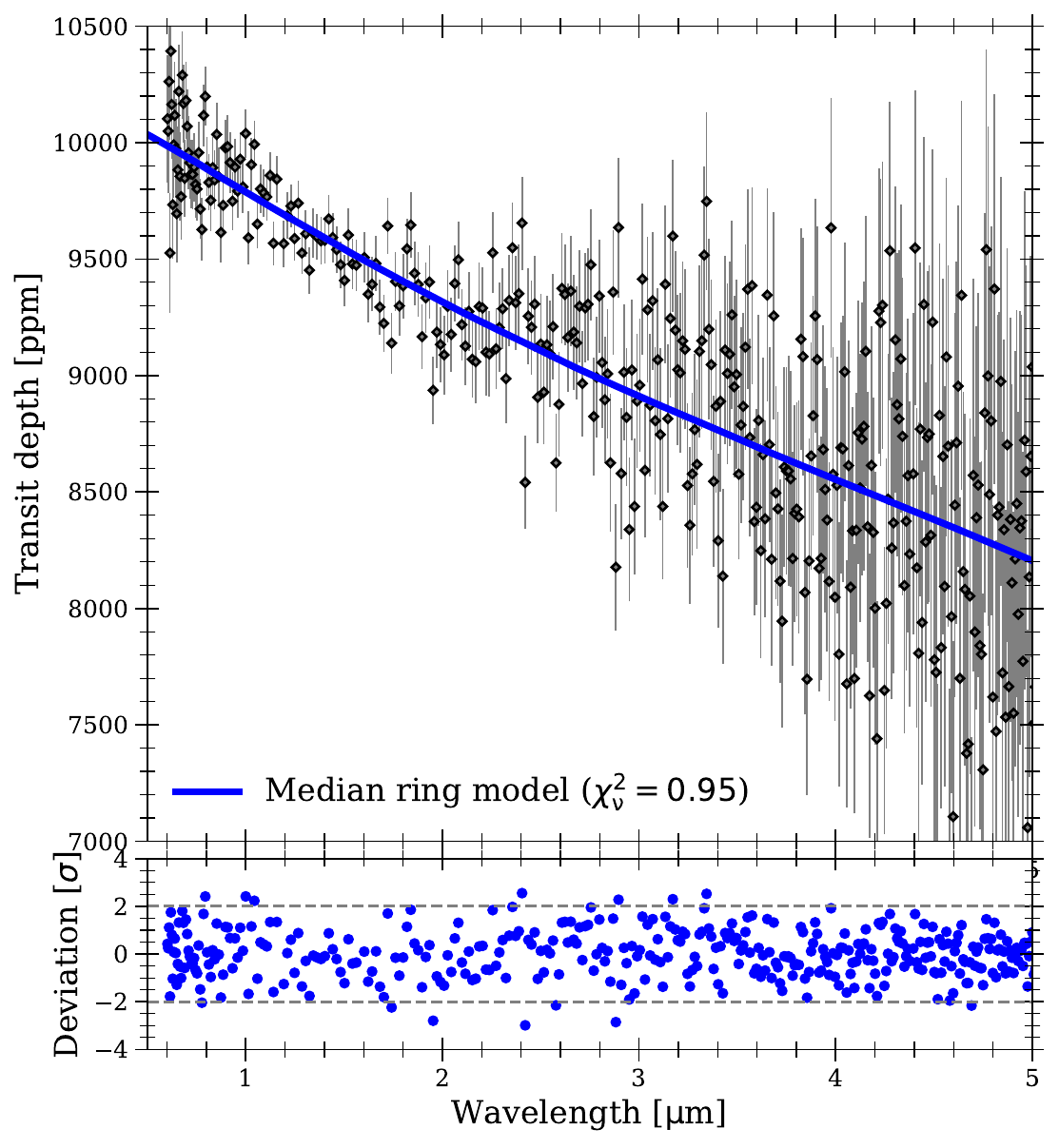}
    \includegraphics[width=0.51\linewidth]{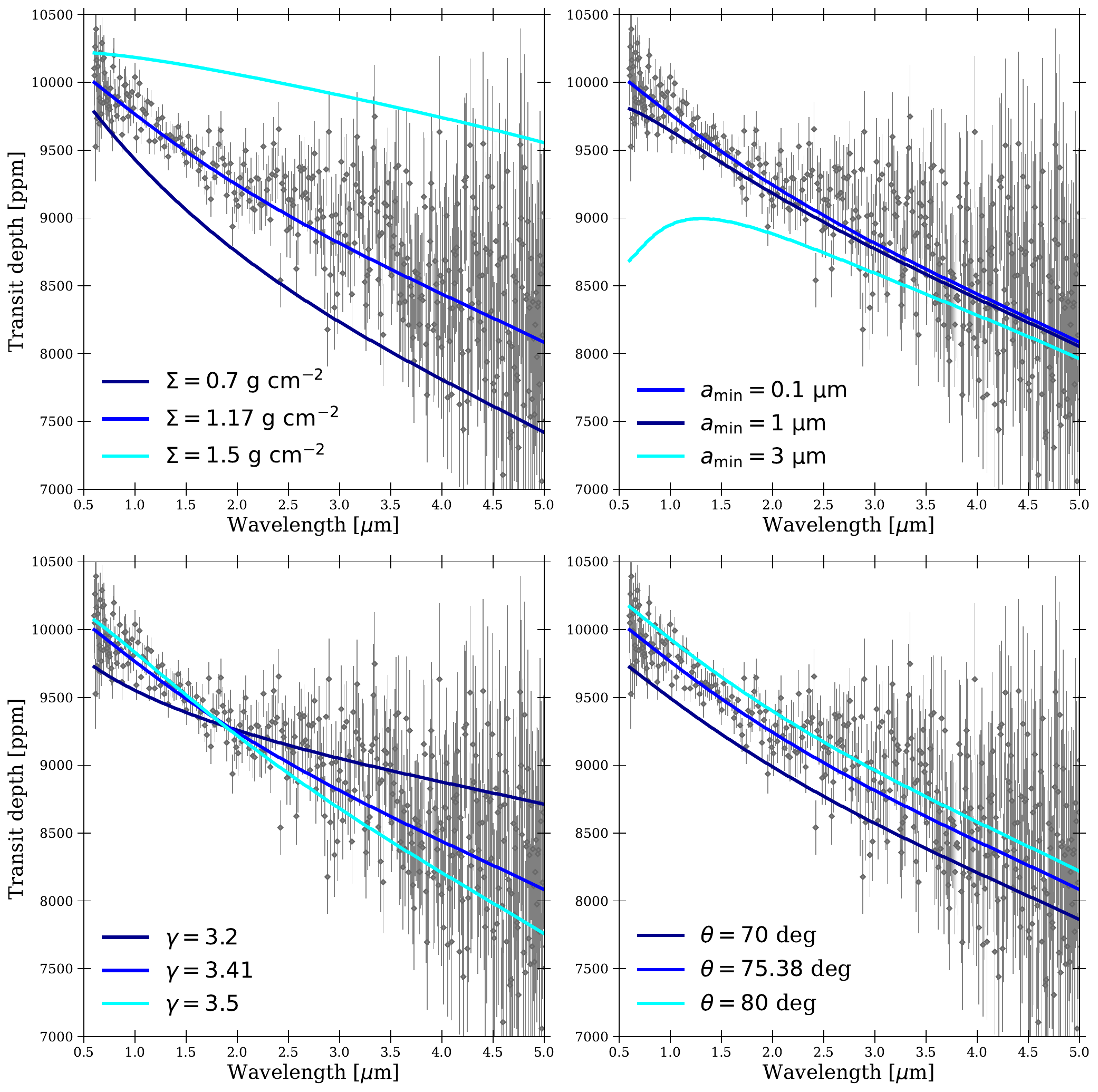}
    \caption{Median ring spectrum (left top) and the model deviation from each data point (left bottom). The right four panels demonstrate how the median spectrum gets affected by the perturbation on ring column mass density $\Sigma$, minimum particle size $a_{\rm min}$, power-law index of the size distribution $\gamma$, and ring obliquity $\theta$.}
    \label{fig:ringspec}
\end{figure*}

Circumplanetary rings have also been proposed as an explanation for the anomalously inflated radii of super-puffs \citep[e.g.,][]{Zuluaga+15,piro.rings}. However, an optically thick ring is expected to produce a relatively flat transmission spectrum \citep{OhnoFortney22_ringspec,alam.hip}, which is distinct from the observed spectrum of Kepler-51d. Thus, our observation suggests that Kepler-51d unlikely possess a massive opaque ring; otherwise, we would observe a flat spectrum rather than a spectral slope.

Nonetheless, Kepler-51d might still have an optically-thin circumplanetary ring that could produce a steep spectral slope if it contains abundant small particles \citep{Ohno22_ringretrieval}. Here, we apply the ring retrieval model of \citet{Ohno22_ringretrieval} to assess whether the tenuous ring could explain the observed spectrum of Kepler-51d.
The ring outer edge is assumed to be at the planet's Roche radius \citep{Schlichting&Chang11_ring,piro.rings}.
We make particle porosity a free parameter that is needed to explain the radius of Kepler-51d \citep{piro.rings}.
We use the open-source Mie theory code \texttt{PyMieScatt} \citep{Sumlin+18_PyMieScatt} to compute the ring optical depth, where the effective medium theory with the Bruggeman mixing rule \citep[e.g.,][]{bohren2008} is adopted to take into account the effects of porosity on the particle opacity.
A power-law size distribution is assumed for ring particles, as seen in circumplanetary rings in the Solar System \citep[e.g.,][]{Cuzzi2009}.
We adopt the astronomical silicate \citep{Draine+03_astrosil} as a refractive index of ring particles.
In total, we vary the column mass density of ring particles $\Sigma$, minimum particle radius $a_{\rm min}$, power-law index of the size distribution of ring particles $\gamma$, particle porosity $\mathcal{P}$, and ring's obliquity as free parameters, where the maximum particle radius is fixed to $a_{\rm max}=10~{\rm m}$ which is the value of a Saturnian ring \citep{Cuzzi2009}.
\texttt{PyMultinest} \citep{Buchner+14_pymultinest} is used to obtain the posterior distributions for each parameter.
We treat a ring-free planet as a rigid sphere with a radius of $2R_{\rm \oplus}$ and ignore atmospheric features, as our observed spectrum is almost completely featureless (Section \ref{sec:transmission_spectrum}).

The hypothetical ring model could still well explain the observed spectrum with the reduced chi-squared value of $\chi^2_{\rm \nu}=0.95$, as shown in Figure \ref{fig:ringspec}.
The ring optical depth decreases from $\approx 3$ at $1~{\rm {\mu}m}$ to $\approx1.5$ at $5~{\rm {\mu}m}$ in the median ring model.
The ring's surface density would be $\approx1~{\rm g~cm^{-2}}$ to yield a proper optical depth, although it can be degenerate with the maximum size of the ring particles.
The minimum size of the ring particles should be $<1~{\rm {\mu}m}$; otherwise, the ring could nor explain the optical spectral slope seen in the data.
The ring model could explain the observed spectral slope if the size distribution obeyed $d\Sigma/da\approx Ca^{-3.4}$, which is close to the outcome expected for the collision cascade with size-independent collision velocity \citep{Tanaka+96,Brillaiantov+15}.
The ring would have a large obliquity of $\sim75~{\rm deg}$ (i.e., $i\sim15$ deg in our setup) to ensure the large observed radius is not inconsistent with the shape of the observed light curve \citep[$i<27~{\rm deg}$,][]{lammers.and.winn}.
The median model also requires a high particle porosity of $\approx85\%$, which is similar to Saturnian ring particles \citep{Zhang2019_ring_porosity}, to ensure a large enough ring \citep{piro.rings}.
Although this ring model requires specific ring properties, we cannot rule out this possibility solely from the data.

The present analysis suggests the need of submicron ring particles; however, they can readily spiral to the planet through Pointyng-Robertson drag \citep{Schlichting&Chang11_ring}.
In the limit of optically thin ring, the orbital decay time of individual particle is given by \citep{Schlichting&Chang11_ring}
\begin{eqnarray}\label{eq:t_PR}
    t_{\rm PR}&\sim& \frac{8\rho_{\rm p}a_{\rm p}c^2}{3(L/4\pi a^2)Q_{\rm PR}(5+\cos^2{\theta})}\approx\frac{2\rho_{\rm p}a_{\rm p}c^2}{15\sigma T_{\rm eq}^4Q_{\rm PR}}\\
    \nonumber
    &\approx&\frac{400~{\rm yr}}{Q_{\rm PR}}\left( \frac{a_{\rm p}}{1~{\rm {\mu}m}}\right)\left( \frac{T_{\rm eq}}{360~{\rm K}}\right)^{-4}\left( \frac{\rho_{\rm p}}{1~{\rm g~cm^{-3}}}\right).
\end{eqnarray}
where $a_{\rm p}$ is the particle radius, $c$ is the speed of light, $L$ is the stellar luminosity, $\theta$ is the inclination of the ring orbital plane with respect to the planetary orbital plane, and $Q_{\rm PR}$ is the radiation pressure efficiency factor \citep{Burns79_PR}.
The lifetime of submicron particles is indeed short.
To assess the actual ring lifetime, however, one needs to take into account the particle size distribution.

Here, we estimate the ring lifetime with a size distribution as follows.
Pointing-Robertson drag quickly removes tiny particles, whereas collisional fragmentation of large particles continuously replenishes such tiny particles. 
Assuming that the collision cascade is fast enough to maintain the power-law size distribution $\mathcal{N}(a_{\rm p})=Ca_{\rm p}^{-\gamma}$ with $C=3(4-\gamma)\Sigma/4\pi\rho_{\rm p}(a_{\rm max}^{4-\gamma}-a_{\rm min}^{4-\gamma})$, where $\mathcal{N}$ is the column number density, the lifetime of an optically-thin ring can be estimated as
\begin{eqnarray}
\nonumber
    t_{\rm life}&\equiv& \Sigma \left|\frac{d\Sigma}{dt}\right|^{-1}\sim \Sigma\left(\int_{a_{\rm min}
}^{a_{\rm max}}\frac{4\pi a_{\rm p}^{3}\rho_{\rm p}}{3}\frac{\mathcal{N}(a_{\rm p})}{t_{\rm PR}(a_{\rm p})}da_{\rm p}\right)^{-1}\\
&\approx&\frac{2\rho_{\rm p}(a_{\rm max}^{4-\gamma}-a_{\rm min}^{4-\gamma})c^2}{15\sigma T_{\rm eq}^4(4-\gamma)}\left(\int_{a_{\rm min}
}^{a_{\rm max}}a_{\rm p}^{2-\gamma}Q_{\rm PR}da_{\rm p}\right)^{-1}
\end{eqnarray}
Since most of ring particles have sizes much larger than the peak wavelength of incident stellar light, we assume $Q_{\rm PR}\sim1$ and then obtain the ring lifetime as
\begin{eqnarray}\label{eq:t_life}
    \nonumber
    t_{\rm life}&\approx& \frac{2\rho_{\rm p}c^2}{15\sigma T_{\rm eq}^4}\frac{(3-\gamma)(a_{\rm max}^{4-\gamma}-a_{\rm min}^{4-\gamma})}{(4-\gamma)(a_{\rm max}^{3-\gamma}-a_{\rm min}^{3-\gamma})}\approx t_{\rm PR}(a_{\rm min})\left( \frac{a_{\rm max}}{a_{\rm min}}\right)^{4-\gamma},
\end{eqnarray}
where we have assumed $3<\gamma<4$ and $a_{\rm max}\gg a_{\rm min}$.
For Kepler-51d, the median ring model infers $a_{\rm min}\sim0.1~{\rm {\mu}m}$, $\rho_{\rm p}\sim0.3~{\rm g~cm^{-3}}$, and $\gamma\sim3.5$, which yields $t_{\rm life}\sim0.1~{\rm Myr}$ for $a_{\rm max}=10~{\rm m}$.
The ring required to explain the observed spectrum has a lifetime much shorter than the age of the Kepler-51 system \citep[500 Myr,][]{libbyroberts2020}.
Thus, we suggest that the hypothetical ring is unlikely the cause of the observed spectrum, unless the ring formed through very recent transient events.

\subsection{Haze Production on Cool Neptunes and Sub-Neptunes}

Early work by \citet{crossfield2017} performed an initial population study of warm and cool Neptunes and sub-Neptune planets with HST/WFC3 observations. Their goal was to search for trends between the strength of the water absorption feature at 1.4 $\mu$m against a variety of planetary and stellar parameters. With a sample size of six planets, \citet{crossfield2017} found two possible linear correlations: (i) the hotter the planet, the larger the feature, or (ii) the lower the atmospheric mean molecular weight, the larger the feature. With the HST/WFC3 featureless spectra of Kepler-51b and -51d, \citet{libbyroberts2020} found strong support for the temperature trend while disfavoring the mean molecular weight one. Additional studies \citep[e.g.][]{yu2021,dymont.haze.trend,brande.trends} have corroborated this trend, even hypothesizing a second-degree function compared to a linear one. Planets appear to be increasingly hazy (featureless) from 700 K to 300 K before becoming increasing clear with decreasing temperature from 300 K to 200 K. While no population study with JWST observations is published to date, individual planet publications with JWST continue to support an overall temperature versus haze dependency. K2-18b and TOI-270d, with temperatures of 215 K \citep{hu2021} and 340 K \citep{toi270.discovery} respectively, demonstrate clear features of water and/or methane and lack of significant hazes \citep{madhusudhan2023k218b,Benneke2024,Holmberg2024}. At 230 K, LHS 1140b does not demonstrate any large features, though \citet{cadieux.lhs1140b} attribute this to a potential secondary atmosphere with a higher mean molecular weight (and therefore decreased scale height) as opposed to high altitude haze production. \citet{damiano.lhs1140b} finds a similar result for LHS 1140b though also include a cloud-deck in their retrievals driven by condensation (as opposed to photochemically produced hazes). The hotter 770 K Neptune, WASP-107b demonstrates a wealth of absorption features \citep{sing.wasp107,Welbanks2024}, and the 920 K Neptune TOI-421b also exhibits multiple water features \citep{Davenport2025_toi421b_haze_free}. GJ 1214b (550 K), GJ 3470b (600 K) and TOI-863c (600 K) all appear to maintain substantial high-altitude haze layers that nearly mute or flatten their respective transmission spectra \citep{gj1214.jwstspectrum,gao2023,Schlawin2024_gj1214b,Ohno2025_gj1214b,beatty.gj3470b,wallack.toi836c}. 

One hypothesis as to why the 300 - 500 K range is prime for haze production in Neptunes/sub-Neptunes is the presence of methane as the dominate carbon molecule at lower pressures \citep{morley2015,kawashima2019b,gao2020} combined with the proximity to the host star for substantial photodissociation \citep{yu2021}.
As observed in our cold Solar System gas giants and Titan, the photodissociation of methane leads to higher order hydrocarbon-based hazes at high-altitudes \citep[][and references therein]{solarsystem.haze}. Sulfur-based hazes (S$_8$-hazes) are also a possibility at these temperatures \citep{gao2017sulfur}. At 350 K, Kepler-51d's low-density, low-surface gravity atmosphere likely possesses the right combination of chemistry for haze production. This, combined with its active and energetic host star, makes Kepler-51d the coolest haze-dominated atmosphere known to date.

\section{Discussion - Stellar Activity}\label{sec:stellar_discussion}

\subsection{Transit Spot-Crossing}
\begin{figure*}
    \centering
    \includegraphics[width=.8\linewidth]{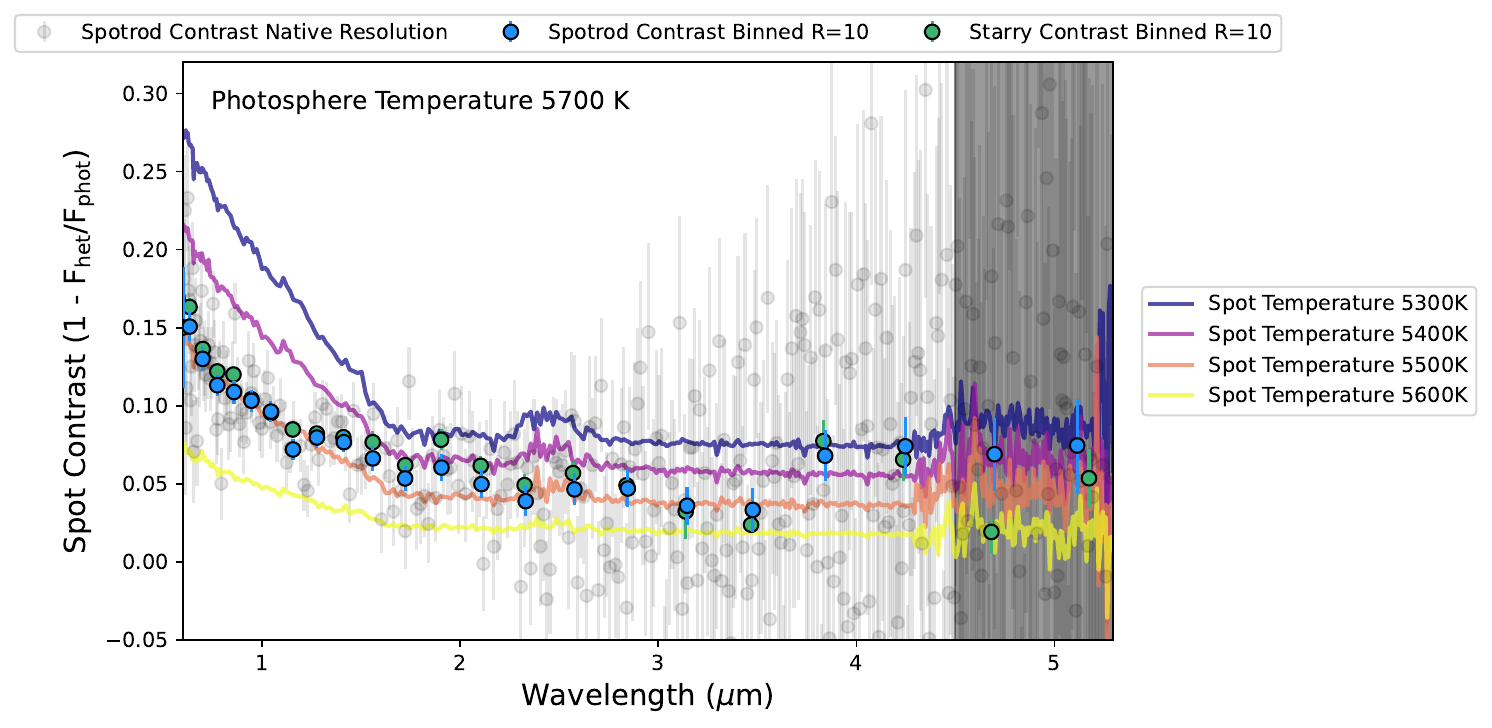}
    \caption{Best-fit spot contrasts from the individual channels (\textit{black}) fit with \texttt{spotrod}, and binned to R=10 for clarity (\textit{blue}). Best-fit spot contrasts derived with \texttt{starry} on R=10 binned spectroscopic light curves are plotted in green. Regardless of model or method (fit then bin versus bin then fit), the spot contrasts from both are statistically the same. Assuming a photosphere temperature of 5700 K, we include a range of PHOENIX-derived spot temperature models for comparison and ignore wavelengths $>$ 4.5 $\mu$m which demonstrate significant scatter in the individual channels. From this analysis, the measured spot contrasts suggest a spot temperature $\Delta$T 200 -- 300 K cooler than the photosphere.}
    \label{fig:spotcontrast}
\end{figure*}

Using the spot contrasts derived from both the native resolution light curves fit with \texttt{spotrod} (Section~\ref{sec:spectroscopiclc}) and the binned spectroscopic light curves fit with \texttt{starry} (Section~\ref{sec:spectroscopiclc_starry}), we derive an approximate spot temperature difference using:

\begin{equation}
    \mathrm{Contrast} = 1 - \frac{F_{spot}}{F_{phot}}
\end{equation}

where F$_{spot}$ and F$_{phot}$ are the expected wavelength dependent flux from the cooler spot and hotter photosphere respectively \citep{spotrod}. In this formalization, which is adopted by \texttt{starry}, a spot temperature similar to the photosphere temperature would yield a contrast approaching 0. We note that the contrast is the opposite in \texttt{spotrod} which uses the fraction of the flux ($\frac{F_{spot}}{F_{phot}}$). In order to compare both analyses consistently, we subtract 1 from the measured contrast of \texttt{spotrod}-derived contrasts and flip the upper and lower uncertainty bounds. The best-fit contrasts are plotted in Figure~\ref{fig:spotcontrast} where we also bin the native resolution contrast to the same resolution as \texttt{starry} for clarity. We adopt the different best spot-configurations derived by \texttt{spotrod} and \texttt{starry} in Section~\ref{sec:spectroscopiclc} and Section~\ref{sec:spectroscopiclc_starry} respectively. Interestingly, we determine statistically the same contrast for the spot regardless of configuration in size and location.

We use \texttt{PHOENIX} stellar models \citep{husser.phoenix} to derive a range of stellar models for different temperatures using 0.05 metalicity and 4.7 log(g) \citep{libbyroberts2020}. Assuming a photosphere temperature of 5800 K \citep{libbyroberts2020}, the derived contrasts from both analyses support a spot temperature between 5400 -- 5500 K. Slightly increasing or decreasing the photosphere temperature leads to different spot temperatures though the models suggest a temperature difference $\Delta$T $\sim$ 200 -- 300 K or a T$_{spot}$/T$_{phot}$ $\sim$0.95.

This spot temperature is $>$ 1000 K hotter than the average temperature of Sunspot umbras which range between 3000 K and 4500 K \citep[e.g.][]{sunspot.temperature}, and could suggest Kepler-51 possesses weaker localized magnetic fields \citep{spottemp.magfields}. Multiple investigations of younger T-Tauri stars have discovered a wide-range of spot temperatures, including multiple young stars with spot temperatures $<$ 1000 K compared to the photosphere \citep[e.g][]{spottemp.youngstars.2009,spottemp.youngstars.2016}. Alternatively, this temperature is similar to the hotter penumbra observed surrounding Sunspots which have temperature differences of $\sim$300 K compared to the photosphere \citep{sunspot.penumbra} The grazing nature of this spot crossing event (Figure~\ref{fig:wlc}) could also indicate that we are probing the spot's penumbra on Kepler-51 and not the cooler umbra portion -- assuming this 500 Myr Sun-like star possesses similar spot structures as the Sun. On the other hand, this bump-like feature could be due to a spot complex, with multiple smaller cool spots and hot faculae. Given the limitations of our spot model requiring a circular spot of a single contrast, we are retrieving the average contrast of this complex. 

Currently, there is very little information regarding spot temperatures on Sun-like stars $<$ 1 Gyr - with this work providing the first direct temperature measurement via spot crossing. Future work of Kepler-51 and similar stars is necessary in order to conclude that these younger stars, while more ``spotty", possess hotter spots.

\subsection{Out of Transit Stellar Spectrum} \label{sec:ootstellarspectrum}

To test the impact of unocculted spots on the transmission spectrum of Kepler-51d, we analyzed the out of transit stellar spectrum. We re-reduced the  \texttt{uncal} JWST data with \texttt{Eureka!} following the procedure outlined in \citet{may.gj1132} which includes flat fielding, throughput correction, and converting the extracted 1D stellar spectrum to flux density as mJy (compared to units of photons). We then median combined all stellar spectra pre-transit and again post-transit. Dividing the pre- and post-transit spectra shows no detectable deviation, indicating that the star experienced no significant changes in activity over the $\sim$14 hours of observation. 

We median combined all out of transit stellar spectra, flagging and removing outliers $>$ 3$\sigma$ from the median in each wavelength channel. We adopted uncertainties following the recommendation of \citet{may.gj1132} who suggested using the standard deviation across each wavelength channel instead of using the standard error. However, as we fit a jitter term to the uncertainty, we found that either form of flux density uncertainty yielded the same results. We used \texttt{pysynphot} \citep{pysynphot2013} function called by \texttt{POSEIDON} \citep{poseidon2} to interpolate over the grid of \texttt{PHOENIX} spectra fitting for a photospheric temperature as well as a scaling factor multiplied to the (R$_s$/distance)$^2$ while holding the stellar metallicity and log(g) constant to 0.05 and 4.7, respectively. We fit the the various parameters using \texttt{emcee} \citep{emcee} with 40 walkers and 10000 steps.

When assuming a homogeneous stellar surface temperature (photosphere only) model, we determined a best-fit photospheric temperature of 5528$^{+69}_{-113}$ K with a $\chi^2_r$ of 4.80. While the photosphere temperature is within 1$\sigma$ of the reported spectroscopic value in \citet{libbyroberts2020}, it does not provide statistically good fit - an obvious outcome given the observed spot crossing event. We note that \citet{may.gj1132} suggests also fitting for the log(g) of the star; however, including this factor yields the same photospheric temperature but a nonphysical log(g) $<$ 3.5. The mass and radius of Kepler-51 is well-known from spectroscopic analysis in multiple sources \citep[e.g.][]{libbyroberts2020,kepler.stars.updated.ages,masuda.kepler51e} so we opted to hold log(g) constant. 

\begin{figure*}
    \centering
    \includegraphics[width=.9\linewidth]{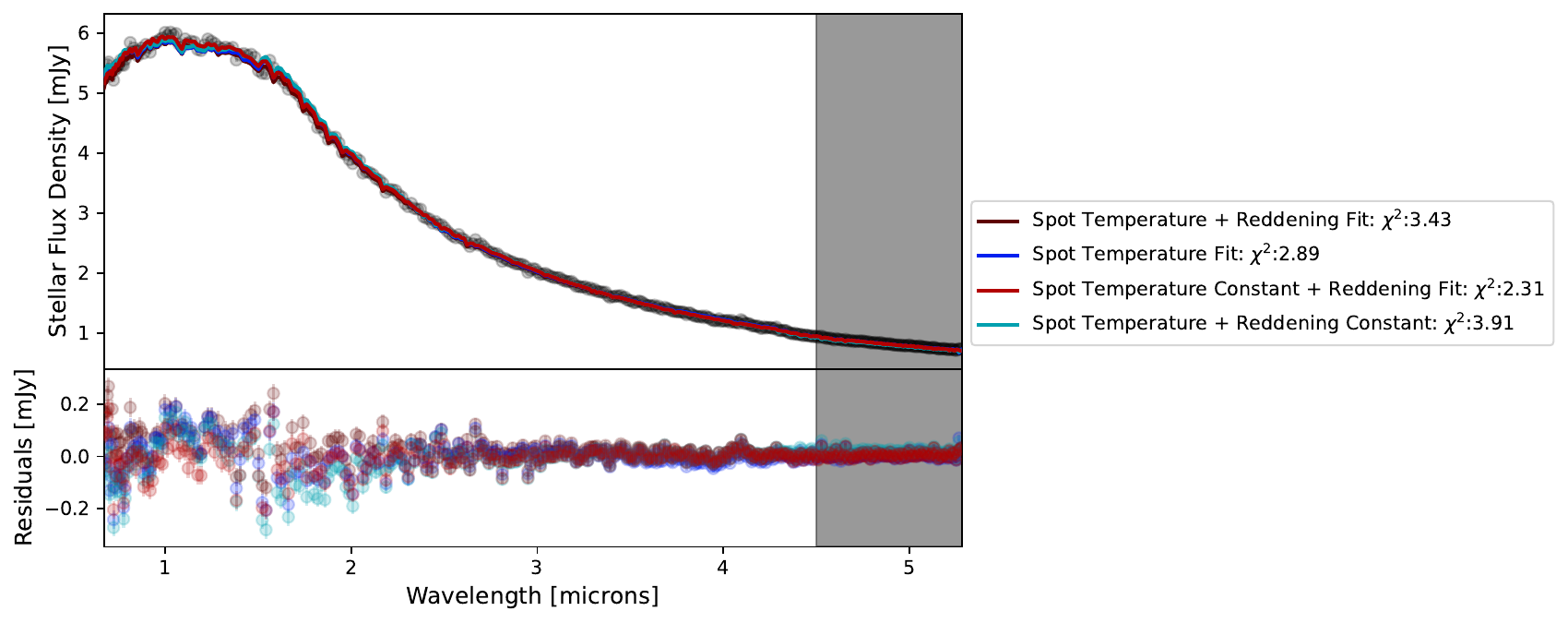}
    \caption{\textit{Top:} Plot of the out of transit stellar flux density along with the best-fit models from four different model assumptions. The dark and light red lines are from models fitting for the overall reddening, while the dark blue points are from the model with no reddening but spot temperature fit and the light blue is a model holding spot temperature and reddening constant to the values derived from the spot crossing and \texttt{dustmaps} respectively. \textit{Bottom:} The residuals color coded based on their respective models. None of the spectral models are able to reproduce the scatter at wavelengths $<$ 3 $\mu$m while all fit the longer wavelengths well.}
    \label{fig:ootspectrum}
\end{figure*}

We perform a similar analysis assuming two temperatures (cooler spot $+$ photosphere) and three temperatures (cooler spot $+$ hotter faculae $+$ photosphere). In each run, we also fit for the spot and faculae coverage fractions, a scale factor multiplied to the (R$_s$/distance)$^2$ correction, and a jitter term multiplied to the flux density uncertainty. We discovered a strong preference for the two temperature model as the best-fit faculae temperature was within 1$\sigma$ of the best-fit photosphere temperature (the two posteriors overlapped) and the faculae coverage fraction was $<$ 1\%. The spot temperature and coverage fraction was identical between the two and three temperature models. We also found the same log(g) issue and again opted to hold this and metallicity constant. We derived a best-fit photosphere and spot temperatures of 5632 $\pm$ 12 K and 3112$^{+117}_{-126}$ K respectively, with a significant spot coverage fraction of 13 $\pm$ 1\% yielding a $\chi^2_r$ of 2.89 (Figure~\ref{fig:ootspectrum}). While this model well-represents wavelengths $>$ 3~$\mu$m, significant scatter remains for the bluer wavelengths. 

Kepler-51 lies $\sim$800 pc from Earth. We therefore expect the stellar light to experience significant `reddening' or extinction. Using \texttt{dustmaps} from \citet{dustmaps.green} combined with the Bayestar map \citep{bayestar.green.2019}, we calculate an expected A$_V$ extinction of 0.20 for Kepler-51. This is non-negligible for the bluer wavelengths of PRISM, and we re-run the same two-temperature model but fitting for an additional factor A$_V$. This scaling parameter is converted into a wavelength dependent factor using \texttt{extinction} package and assuming the \citet{fm07.extinction} scaling relations. We achieve a $\chi^2_r$ of 3.43 with a bi-modal spot temperature with one peak at 3100 K and a second at 5700 K running up against the requirement that the spot temperature be cooler to the photosphere temperature. This bi-modality is correlated with the best-fit extinction parameter though both spot temperatures prefer a slightly larger A$_V$ (0.24 and 0.28 respectively). The two models (with and without reddening included) are indistinguishable at redder wavelengths as expected. Accounting for extinction we derive a hotter best-fit photosphere temperature of 5856 $\pm$ 25 K and a lower spot coverage fraction of 7\% for the cooler spot and unconstrained for the hotter spot temperature. We perform a similar fit, but holding A$_V$ constant at the determined 0.2 value. However, this yielded similar bimodal solutions to those allowing A$_V$ to float with no statistical improvement at shorter wavelengths. 

\begin{figure*}
     \centering
    \includegraphics[width=1\linewidth]{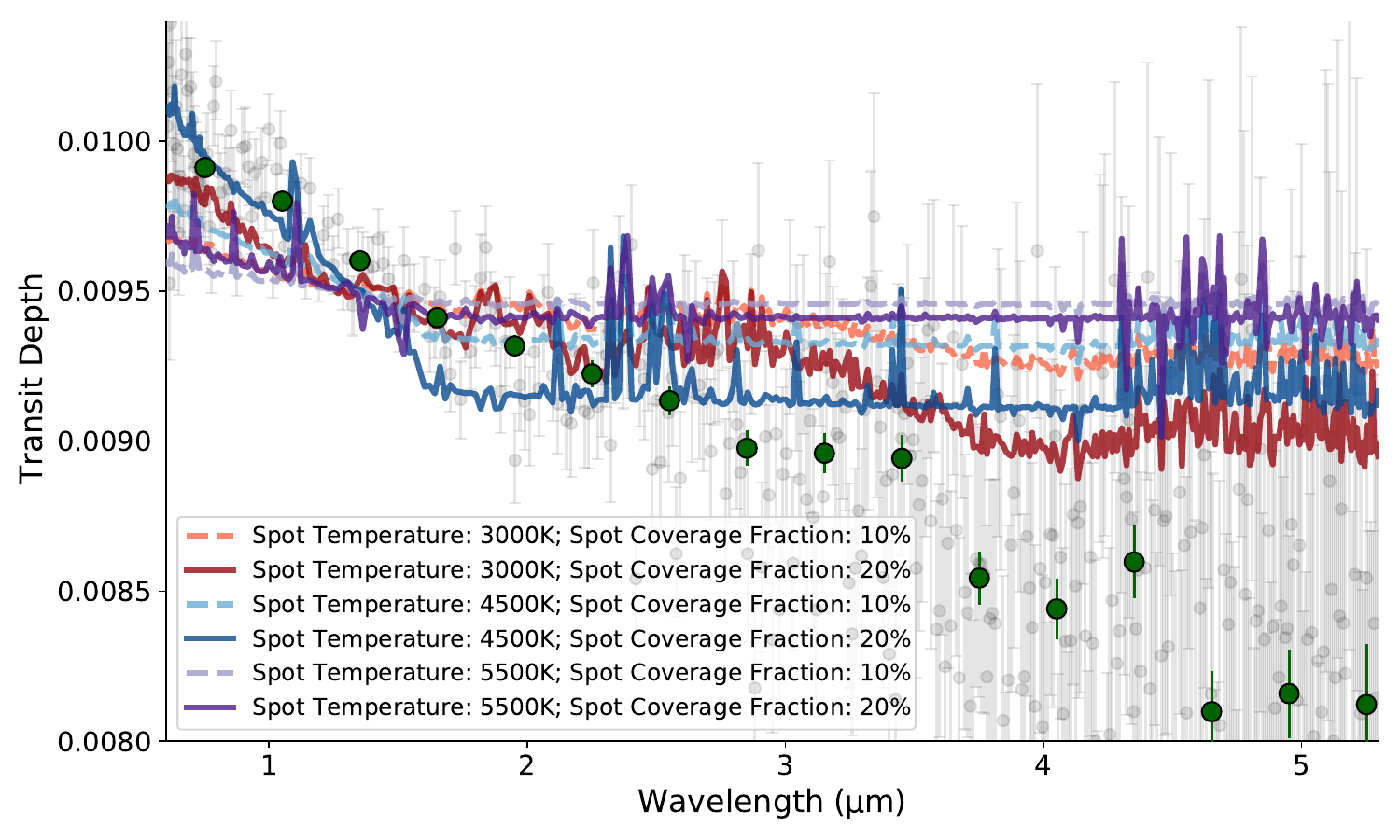}
    \caption{Kepler-51d's transmission spectrum (\textit{gray}) and binned spectrum (\textit{green}) plotted against different simulated transmission spectra assuming only stellar contamination. Each model assumes a photosphere temperature of 5800 K with a varying spot temperature and spot coverage fractions. Spot coverage fractions $<$5\% yielded a nearly flat line or no stellar contamination impact. While cold spots at large coverage fractions could reproduce the bluer $<$2 $\mu$m wavelength slope, no spot temperature and coverage fraction can explain the observed $>$2 $\mu$m slope. Therefore, the transmission spectrum must be due to planetary atmospheric absorption.}
    \label{fig:stellarcontaminationspectrum}
\end{figure*}

All best-fit out-of-transit models demonstrate significant scatter at bluer wavelengths (Figure~\ref{fig:ootspectrum}) while also fitting the data with vastly different stellar parameter values with similar results. It is unclear whether this is due to instrumental effects, issues introduced during the data reduction, an outcome of the low-resolution PRISM, or a limitation in the stellar \texttt{PHOENIX} models for the younger and more active Kepler-51 ($<$ 1 Gyr). We therefore caution in over-interpreting the results of this investigation. That said, none of these models with differing spot temperatures and coverage fractions can reproduce the overall slope observed in the transmission spectrum of Kepler-51d (Figure~\ref{fig:stellarcontaminationspectrum}). Moreover, the cooler spot models ($<$ 3500 K) also introduce water features into the spectrum, which we do not retrieve. The hotter spots with smaller coverage fraction has no impact on the transmission spectrum. We therefore concluded that the transmission spectrum of Kepler-51d is unaffected by unocculted stellar activity to detectable levels.

\section{Conclusion} \label{sec:conclusion}

The Kepler-51 system, comprised of a 500 Myr Sun-like star hosting three low-density transiting super-puff planets and one low-mass non-transiting planet, remains unique across all known systems to date. How this system formed and evolved depends on our knowledge of the composition and structure of the individual planets. Observations of Kepler-51b and Kepler-51d with HST/WFC3 yielded surprisingly featureless spectra \citep{libbyroberts2020} that are explainable with various hypotheses ranging from high-altitude hazes and escaping atmospheres \citep{kawashima2019a,wang.and.dai,libbyroberts2020,gao.and.zhang,ohno2021sp} to tilted planetary rings \citep{akinsanmi.rings,piro.rings,OhnoFortney22_ringspec}. Observations with JWST's larger aperture and longer wavelength coverage than HST should test these hypotheses. We observed a single transit of Kepler-51d, the furthest transiting planet from its star. Using the NIRSpec/PRISM, we obtained a near-featureless, strongly sloped planetary transmission spectrum from 0.6 -- 5.3 $\mu$m. We retrieved a tentative 2.2 $\sigma$ detection of methane on the order of ppb, which we conclude is due to possible upwelling from the interior and not representative of the total methane abundance. The sloped shape of the transmission model is explainable via both retrievals and forward models as a substantial high-altitude haze layer spanning pressures from 1-100 $\mu$bars on top of an extended H/He atmosphere. Forward models suggest sub-micron particle sizes akin to the tholin haze particles in Titan's haze layer while retrievals suggest a particle sizes a few times larger. However, both the forward models and retrievals find haze compositions of tholin, soot, and sulfur are equally likely to reproduce the transmission spectrum. 

We also explore the possibility that the transmission spectrum is due to a thin, tilted ring system of porous materials. Our sloped spectrum convincingly rules out a massive, opaque ring as the cause of the large radius of Kepler-51d, whereas we find that an optically-thin ring with abundant submicron particles can still reproduce Kepler-51d's sloped spectrum. However, the smallest ring particle should experience significant Pointing-Robertson drag, requiring the larger particles to constantly replenish the smaller particles through collisions. 
From a simple analytical argument, we estimated the lifetime of the hypothetical ring to be $\sim$0.1 Myr.
Given this timing is substantially smaller than the system lifetime, a stable ring is an unlikely explanation for Kepler-51d. However, we cannot rule-out the possibility this is a transient event created by a recent collision(s).

As Kepler-51 is an active star, we test the possibility that some, if not all, of the transmission spectrum is due to stellar contamination commonly observed in other sub-Neptune JWST spectra \citep[e.g.][]{moran.stellarcontamination,trappist1b.contamination}. All of these observations are of cooler M-dwarf stellar hosts while Kepler-51 is a hotter G-dwarf. We leverage a star spot crossing event mid-transit to measure a wavelength dependent spot contrast -- finding a $\Delta$T $\sim$300 K between the photosphere and spot. Based on this analysis, we conclude Kepler-51 possesses spots $>$ 1000 K hotter than those on our own Sun. Alternatively, it is possible that the spot feature is comprised of a hotter spot penumbra or a spot complex of hot and cold spots. Out-of-transit stellar spectra are unable to constrain the spot temperature or the coverage fraction. Regardless, we find that any best-fit spot/photosphere temperature/coverage fraction cannot explain the large slope observed from 0.6 -- 5.3 $\mu$m. We therefore conclude that Kepler-51d's inflated radius is likely due to a large and extended H/He atmosphere with a substantial haze layer comprising at least two orders of magnitude in pressure, albeit, we cannot rule-out the possibility of a short-lived ring system. Future observations of the other super-puff planets in the Kepler-51 system with JWST could provide additional insights into how these planets (Kepler-51d included) formed and whether they all possess a substantial haze layer. For now, Kepler-51d is the only known planet with a featureless sloped JWST transmission spectrum spanning 0.6 -- 5.3 $\mu$m.

\begin{acknowledgments}

We are extremely grateful to Weston Eck, Elena Manjavaca, and the staff at Space Telescope Science Institute for scheduling this observation and accommodating our last minute instrumental set-up requests. We thank Eve Lee for many insightful discussions on super-puff formation, Patrick McCreery for updating Kepler-51 stellar abundances, Andrew Mann for advice on modeling out-of-transit stellar spectra, and Allison Youngblood and David Wilson for sharing their UV stellar constraints for Kepler-51. We also wish to thank Heather Knutson for her thoughtful feedback on the JWST 2571 proposal which contributed to its selection and the resulting observation.

This work is based on observations made with the NASA/ESA/CSA James Webb Space Telescope. The data were obtained from the Mikulski Archive for Space Telescopes at the Space Telescope Science Institute, which is operated by the Association of Universities for Research in Astronomy, Inc., under NASA contract NAS 5-03127 for JWST. These observations are associated with program 2571.

The authors of this work recognize the Penn State Institute for Computational and Data Sciences (RRID:SCR\_025154) for providing access to computational research infrastructure within the Roar Core Facility (RRID: SCR\_026424).

The Center for Exoplanets and Habitable Worlds is supported by the Pennsylvania State University, the Eberly College of Science, and the Pennsylvania Space Grant Consortium. 

YK acknowledges support from JSPS KAKENHI Grant Numbers 21K13984, 22H05150, and 23H01224. Part of this research was carried out at the Jet Propulsion Laboratory, California Institute of Technology, under a contract with the National Aeronautics and Space Administration (80NM0018D0004).
CIC acknowledges support by NASA Headquarters through an appointment to the NASA Postdoctoral Program at the Goddard Space Flight Center, administered by ORAU through a contract with NASA. 

\end{acknowledgments}
\begin{contribution}

JLR:
Led the overall project, including performing data reductions and analyses, supervising atmospheric modeling and retrievals, and drafting and submitting the manuscript.

ABA:
Performed an independent data reduction and fitting for comparison, and contributed to the writing of the manuscript.

ZBT:
Served as co-PI on the JWST proposal, provided the chromatic fitting package, and provided feedback on the manuscript.

CC:
Performed an independent data reduction and fitting for comparison, including modifying Eureka! for the spotrod implementation, and contributed to manuscript writing.

YC:
Performed PLATON retrievals of the transmission spectrum, contributed to haze-related scientific discussions, and participated in manuscript writing.

RH:
Developed ExoTR, advised ABA and AT, supervised retrieval efforts, contributed to haze discussions, and provided manuscript feedback.

YK:
Developed forward models for comparison with the observed spectrum and retrievals, contributed to haze discussions, and participated in writing the manuscript.

Catriona Murray:
Performed an independent analysis using starry, investigated starspot properties, contributed to discussions on stellar activity, and participated in manuscript writing.

KO:
Conducted a detailed investigation into a circumplanetary ring scenario, contributed to haze discussions, and participated in manuscript writing.

AT:
Performed ExoTR retrievals of the transmission spectrum, contributed to haze discussions, and participated in manuscript writing.

SM:
Advised JLR, helped oversee the entire project, and provided feedback on the manuscript.

KM:
Provided an updated planetary mass for Kepler-51d used in atmospheric retrievals and reviewed the manuscript.

LH:
Contributed to discussions on stellar activity and reviewed the manuscript.

Caroline Morley:
Contributed to haze-related discussions and reviewed the manuscript.

GF:
Assisted with the Eureka! data reduction of the transmission spectrum and reviewed the manuscript.

PG:
Provided updated sulfur opacities for haze modeling and reviewed the manuscript.

KS:
Assisted with the Eureka! data reduction of the transmission spectrum and reviewed the manuscript.


\end{contribution}

%

\facilities{JWST(NIRSpec-PRISM)}

\software{  
\texttt{astropy} \citep{AstropyCollaboration2018},
\texttt{chromatic} \citep{zach_berta-thompson_zkbtchromatic_2025},
\texttt{chromatic\_fitting} \citep{catriona_murray_chromatic_fitting_2025},
\texttt{dynesty} \citep{dynesty1},
\texttt{Eureka!} \citep{Bell2022},
\texttt{ExoTR}
\texttt{ExoTiC-JEDI} \citep{Alderson2022},
\texttt{ExoTiC-LD} \citep{exoticld},
\texttt{jwst} \citep{bushouse2023},
\texttt{matplotlib} \citep{Hunter2007},
\texttt{numpy} \citep{vanderWalt2011},
\texttt{pandas} \citep{McKinney2010},
\texttt{PLATON} \citep{platon1,platon2},
\texttt{POSEIDON} \citep{MacDonald2017},
\texttt{scipy} \citep{Virtanen2020},
\texttt{spotrod} \citep{spotrod},
\texttt{starry} \citep{starry2021}
          }


\bibliography{main}{}
\bibliographystyle{aasjournalv7}



\end{document}

%% file: bestfit_table.tex
\begin{table*}[]
\resizebox{\textwidth}{!}{%
\begin{tabular}{lccll}
\hline
\multicolumn{2}{|c}{Parameter}                                & \textbf{Eureka!}                                & \multicolumn{1}{c}{\textbf{Starry}} & \multicolumn{1}{c|}{\textbf{ExoTiC-JEDI}} \\ \hline
R$_p$/R$_s$                   & {[}unitless{]}                & \textbf{0.09682 $\pm$ 0.00051}                  &  $\mathbf{0.09835^{+0.00010}_{-0.00009}}$                          &     $\mathbf{0.0984 \pm 0.0003}$                             \\
Mid-Transit Time (T$_0$)      & {[}BJD$_\mathrm{TDB}${]}      & \textbf{60121.347283 $\pm$ 0.000056}            & $\mathbf{60121.347269^{+0.000037}_{-0.000036}}$                        &     $\mathbf{60121.34732_{-0.00004}^{+0.00005}}$                             \\
a/R$_s$                       & {[}unitless{]}                & $\mathbf{124.16 \pm 0.37}$                     & $\mathbf{124.83 \pm 0.34}$ &   $\mathbf{124.16_{-0.15}^{+0.26}}$                               \\
Inclination (i)               & {[}degrees{]}                 & \textbf{89.870 $\pm$ 0.005}                     & \textbf{89.882 $\pm$ 0.005}                           &        $\mathbf{89.868_{-0.002}^{+0.004}}$                          \\
Period                        & {[}days{]}                    & \textbf{130.18 (fixed)}                         &  \textbf{130.1845 (fixed)}                          &     $\mathbf{130.185_{-0.003}^{+0.007}}$                             \\
Eccentricity                  & {[}unitless{]}                & \textbf{0.0 (fixed)}                            &  \textbf{0.0 (fixed)}                          &   \textbf{0.0 (fixed)}                                 \\
Limb-Darkening (q$_1$, q$_2$) & {[}unitless{]}                & \textbf{(0.131 $\pm$ 0.013, 0.311 $\pm$ 0.047)} & $\mathbf{(0.200^{+0.003}_{-0.002}, 0.393^{+0.408}_{-0.149})}$                           &    $\mathbf{(0.116_{-0.007}^{+0.008},0.367_{-0.032}^{+0.028})}$                              \\
Spot Contrast$^a$             & {[}unitless{]}                & \textbf{0.089 $\pm$ 0.015}                      & $\mathbf{0.087 ^{+0.005}_{-0.008}}$                           &           $\mathbf{0.096 \pm 0.009}$                       \\
Main Spot Radius$^b$          & {[}R$_\mathrm{spot}$/R$_s${]} & \textbf{0.300 $\pm$ 0.031}                      &  $\mathbf{0.207 ^{+0.006}_{-0.017}}$                         &           $\mathbf{0.31 \pm 0.02}$                      \\
Main Spot X-position          & {[}unitless{]}                & \textbf{0.130 $\pm$ 0.005}                      &  \textbf{0.114 $\pm$ 0.003}                          &                       $\mathbf{0.126 \pm 0.004}$         \\
Main Spot Y-position          & {[}unitless{]}                & \textbf{-0.505 $\pm$ 0.044}                     & $\mathbf{-0.145 ^{+0.053}_{-0.005}}$                        &         $\mathbf{-0.52_{-0.02}^{+0.03}}$                       \\
Secondary Spot Radius$^c$     & {[}R$_\mathrm{spot}$/R$_s${]} & \textbf{0.620 $\pm$ 0.115}                      & $\mathbf{0.217 ^{+0.037}_{-0.027}} $                         &         $\mathbf{0.73_{-0.11}^{+0.08}}$                         \\
Secondary Spot X-position     & {[}unitless{]}                & \textbf{-0.592 $\pm$ 0.034}                     &  $\mathbf{-0.455^{+0.020}_{-0.018}} $                        &             $\mathbf{-0.56 \pm 0.04}$                      \\
Secondary Spot Y-position     & {[}unitless{]}                & \textbf{0.417 $\pm$ 0.135}                      &   $\mathbf{-0.427^{+0.029}_{-0.032}}$                         &                $\mathbf{0.5 \pm 0.1}$                   
                           
\end{tabular}}
\caption{$^a$Spot contrast for \texttt{spotrod} were converted to \texttt{starry} definition which assumes a spot contrast of 0 is equal to the photosphere and a contrast of 1 is completely dark; $^b$ Main spot values were used for the spectroscopic light curves as well; $^c$ Secondary spot is a small grazing spot that improves the overall white light curve fit but does not impact the spectroscopic light curve fits; The best-fit Eureka! values were assumed for the rest of this work.}
\end{table*}